\documentclass[journal=jacsat,manuscript=article]{achemso}

\usepackage[version=3]{mhchem} 
\usepackage{bm}
\usepackage{multirow}
\usepackage{dcolumn}
\usepackage{graphicx}

\author{Marwa H. Farag}
\email{mfarag3@ford.com}
\author{Joydip Ghosh}
\affiliation[]
{Research and Advanced Engineering, Ford Motor Co., Dearborn, MI 48124, USA}

\title[An \textsf{achemso} demo]
  {Towards the simulation of transition-metal oxides of the cathode battery materials using VQE methods}

\abbreviations{}

\begin{document}




\begin{abstract}

Variational quantum eigensolver (VQE) is a hybrid quantum-classical technique that leverages noisy intermediate scale quantum (NISQ) hardware to obtain the minimum eigenvalue of a model Hamiltonian. VQE has so far been used to simulate condensed matter systems as well as quantum chemistry of small molecules. In this work, we employ VQE methods to obtain the ground-state energy of LiCoO$_2$, a candidate transition metal oxide used for battery cathodes. We simulate Li$_2$Co$_2$O$_4$ and Co$_2$O$_4$ gas-phase models, which represent the lithiated and delithiated states during the discharge and the charge of the Li-ion battery, respectively. Computations are performed using a statevector simulator with a single reference state for three different trial wavefunctions: unitary coupled-cluster singles and doubles (UCCSD), unitary coupled-cluster generalized singles and doubles (UCCGSD) and k-unitary pair coupled-cluster generalized singles and doubles (k-UpCCGSD). The resources in terms of circuit depth, two-qubit entangling gates and wavefunction parameters are analyzed. We find that the k-UpCCGSD with k=5 produces results similar to UCCSD but at a lower cost. Finally, the performance of VQE methods is benchmarked against the classical wavefunction-based methods, such as coupled-cluster singles and doubles (CCSD) and complete active space configuration interaction (CASCI). Our results show that VQE methods quantitatively agree with the results obtained from CCSD. However, the comparison against the CASCI results clearly suggests that advanced trial wavefunctions are likely necessary to capture the multi-reference characteristics as well as the correlations emerging from high-level electronic excitations.

\end{abstract}

\section{Introduction}
Li-ion batteries (LIBs) are chemical energy storage devices, and they are the dominant propulsive energy source for electric vehicles (EVs)\cite{Aurbach11, Isaacs12, Xu14, Li20, Wu19, Winter04, Manthiram17, Whittingham04, Park13,Ceder21, Yushin15, Aminabhavi18, Hu20, Lu20}. Given the accelerated rate of EVs market penetration globally, it is critical that the energy source that will power these EVs is well understood in order to meet global demand.  While LIBs have been the front runner in leading this battery revolution, it is clear that further progress in battery chemistry is needed to not only improve the battery range but also do so  under all performance, comfort and user experience requirements as they are afforded by gas powered vehicles.  Additionally,  it is also clear that alternate battery chemistry variants are critical in order to balance the expected demand and the availability of critical battery materials.   


To further advance the battery technology, we need to improve the battery materials in terms of their energy density, power density, life-cycle, safety, cost, and quite importantly  recyclability\cite{Augustyn20,Tong17,Sun18,Nam14,Cui22}. Computational chemistry can provide insights about the charge/discharge mechanisms, electrochemical and thermal stability, structural phase transition, and surface behavior\cite{Major20,Ong20,Fisher14,Vegg16,DOMPABLO13,Ceder16,Ceder18}, and it plays a vital role to find potential materials that can enhance the battery performance and robustness.

To solve the many-body electronic structure problem and find the energy eigenstate and the corresponding energy eigenvalue, it is essential to choose a method that allows the best trade-off between accuracy and efficiency\cite{Aspuru:13}. This is, however, very challenging for simulations on classical processors because exact method such as full configuration interaction (FCI) \cite{Shepard12} scales factorially $\mathcal{O}(N!)$, while coupled-cluster methods such as CCSD and CCSD(T)\cite{Barlett12} scale as $\mathcal{O}(N^6)$ and $\mathcal{O}(N^7)$, respectively, where $N$ is the number of orbitals\cite{Szabo96,Olsen14}. The unfavorable scaling of these methods makes their application intractable for large systems, such as solid state materials.

Current state-of-the-art methods employ density functional theory (DFT) to simulate solid-state materials\cite{Martin04,Head17,Burke08}. This method scales as $\mathcal{O}(N^3)$ on classical computers, which makes it practical on today's supercomputers. However, DFT has some fundamental limitations\cite{Cremer02,Marzari06,Grimme17,Wolverton15,Grimme06,Yang08}. For example, it is known that DFT fails, even qualitatively, for systems with strong electron correlation\cite{Rubio12,Sadovskii13,Pacchioni15,Pickard14}. Another limitation of the DFT is the self-interaction energy error\cite{Yang08}. That makes its application to certain classes of systems challenging, such as systems with transition metals\cite{Harrison00,Ryde07}. Transition metals are one of the main component in the cathode battery materials. Previous theoretical works\cite{Major18,Perdew85} employed different DFT functionals to predict the band structure of the cathode battery materials. It has been shown that DFT functionals, such as PBE and SCAN \cite{Major18,Perdew85}, cannot reproduce the experimental values of the band gap for some use cases. For example, the reported experimental band gap for the LiMnO$_2$ is 1.64\cite{Cho15} eV, while the PBE and SCAN functionals\cite{Major18} give a band gap of 0.92 and 1.19 eV, respectively. Analyzing the band gap plays an important role to assess the electronic conductivity of the materials. PBE and SCAN functionals were also employed to estimate the voltage\cite{Ceder10,Ceder16a,Major18}. Results have shown that for PBE functional, the voltage profile of LiMnO$_2$ ranges from 2.4-3.2 V and for SCAN functional, the profile ranges from 2.9-3.4 V\cite{Major18}. On the other hand, the experimental voltage profile ranges from 2.0-4.6 V\cite{Gitzendanner99}. Estimating the voltage of the electrodes accurately is essential for understanding the battery stability. It is known that for LIB, the voltage of the electrode must be within the voltage window of the electrolyte. Although treating the self-interaction energy error by using the DFT+U approach\cite{Gironcoli05,Marzari10} shows better agreement with the experimental value of both the band gap and voltage\cite{Major18}, the adjustable U parameter is a species-specific, i.e., the U parameter needs to be recalculated for each material and for every oxidation state for a given material\cite{Ceder04}. Hybrid functionals can, also, be used to reduce the self-interaction energy error by evaluating the exchange energy partly using Hartree Fock method which comes at a computational overhead\cite{Barone99,Bell13}. However, as shown in previous theoretical works, hybrid DFT functionals do not always work well to compute the properties of transition metal systems\cite{Ryde07,Schaefer99,Illas14}. Furthermore, the long range characteristics of the electrostatic potential makes it computationally demanding to evaluate the exact exchange energy in periodic calculations\cite{Gordon08} and, therefore, only short range contribution is usually considered\cite{Ernzerhof03,Bell13,Ceder10}. Another issue with the DFT method is the non-covalent interactions\cite{Grimme17,Gordon08b} which can be corrected using functionals such as DFT+V\cite{Marzari19} and OPTPBE-VDW\cite{Michaelides10}. 

Accurate and universal tools are needed to simulate battery materials and guide the development of the next generation LIBs, in particular when the experimental results are  unavailable. Quantum computing is an emerging new computational paradigm for a wide class of problems that are intractable to existing high performance computers\cite{Nielsen10}. Quantum chemistry simulation, such as the electronic structure problem of molecules and materials, is one of the most promising near-term applications of quantum computers\cite{Aspuru19,Aspuru:20,Narang21,Chan20}. It is expected that quantum computing will allow accurate simulation of chemical systems for industry-relevant length scales. There are two categories of algorithms typically used to perform electronic structure calculations on quantum computers in order to find the ground-state wavefunction and the corresponding energy eigenvalue of a given Hamiltonian. The first algorithm is the quantum phase estimation (QPE)\cite{Gordon05}, which requires quantum error correction and, therefore, is only amenable to a fault-tolerant quantum computer (FTQC). The second one is the variational quantum eigensolver (VQE)\cite{Aspuru:22} that can be implemented on the near-term quantum devices, such as noisy intermediate scale quantum (NISQ) hardware. 

A recent theoretical work\cite{Arrazola22} developed a quantum algorithm based on QPE to find the ground-state energy for cathode battery materials and estimate the cell voltage, ionic mobility, and thermal stability. It was reported that under the assumption of single reference state (Hartree Fock approximation), thousands of logical qubits and trillions of logical gates are required to execute QPE. Another theoretical work\cite{Roberts22} estimated the resources required to simulate electrolyte molecules of LIB on a  photonic FTQC. The authors of this work suggest architectural and algorithmic techniques to allow use of multiple ``magic state fraction'' to produce magic state in parallel. The authors found that for PF$^-_6$ using cc-pVDZ basis a total of 16382 logical qubits and $2.32 \times 10^{12}$ T gates are needed. Clearly, the application of these methods will not be possible on the near-term quantum devices.

In this paper, we simulate cathode battery materials using VQE methods and benchmark  its performance against the classical wavefunction-based techniques. We simulate gas-phase models of the cathode battery material LiCoO2. To estimate the performance of the VQE-method, we compare the results obtained from the VQE simulations with those obtained from the complete active space configuration interaction (CASCI), complete active space self-consistent field (CASSCF), coupled-cluster with singles and doubles excitation (CCSD), and the second-order M\o{}ller-Plesset perturbation theory (MP2) on classical computers\cite{Szabo96,Olsen14}. CCSD scales as  $\mathcal{O}(N^6)$ ($N$ is the number of orbitals), while CASCI and CASSCF scales exponentially with the size of the active space orbitals. MP2 scales as  $\mathcal{O}(N^5)$. It is important to note that a legitimate (classical) benchmark of VQE must include some wavefunction approach so that we can compare the chemical accuracies for a given number of orbitals and electrons.

The rest of the paper is organized as follows. First, we discuss the difference between VQE-UCC trial wavefunctions that are employed in this work and the active space Hamiltonian. Then, we show how the gas-phase model was defined to mimic the building block of the crystal structure of the material and how the active space is selected for the subsequent VQE simulations. Afterward, we explain the basic steps of the VQE simulations as well as the details of computational methods. Next, we present the calculated energies and discuss the difference between the calculated energies using different trial wavefunctions with the aim of understanding the performance and the resources needed for simulating battery materials on the near-term quantum devices. Finally, we summarize our finding.

\section{Theory}\label{theory}

VQE is an iterative quantum-classical variational algorithm. Here, the quantum device is used to prepare a parameterized wavefunction $ \vert\Psi(\bm\theta)\rangle $, while the classical device is used to optimize the parameters $(\bm \theta)$ over a relevant cost function. In the electronic structure problem, the quantum device prepare the wavefunction and measures the expectation value of the Hamiltonian ($ \hat{H} $), while the classical device is used to optimize the vector parameters $ (\bm\theta) $ and calculate the energy, $ \langle \Psi(\bm\theta) \vert \hat{H} \vert \Psi(\bm\theta)\rangle $, where $ \hat{H}=\sum_{i} \hat{h}_i $\cite{Aspuru:16}.

The preparation of the wavefunction consists of two steps. The first step is the construction of the reference state $\vert \phi_0 \rangle$ on the classical device, while the second step is the preparation of the trial wavefunction on the quantum computer. For the second step, three different trial wavefunctions based on the unitary coupled-cluster (UCC) ansatz are employed (UCCSD, UCCGSD, and k-UpCCGSD)\cite{Izmaylov22}. The unitary version of coupled cluster (CC) method is suited on quantum computers due to the properties of the applied gate operation. Furthermore, it is worth mentioning that in contrast to the conventional CC on classical computers which is not variational, the UCC on quantum devices is variational.

In the unitary coupled cluster with singles and doubles excitation (UCCSD), the trial wavefunction is written as\cite{Babbush19}
\begin{equation}
    \vert \Psi(\bm\theta) \rangle= e^{\hat{T}{(\bm\theta)} - \hat{T}^{\dagger} (\bm\theta)} \vert \phi_0 \rangle 
\end{equation}
where the singles and doubles cluster operators $(\hat{T_1}(\bm\theta) - \hat{T_1}^{\dagger} (\bm\theta))$, $(\hat{T_2}(\bm\theta) - \hat{T_2}^{\dagger} (\bm\theta))$, respectively, are given by 
\begin{align}
  \hat{T}_1(\bm\theta) - \hat{T}_1^{\dagger} (\bm\theta) &= \sum_{i,m} \theta_i^m \left( \hat{a}_m^{\dagger} \hat{a}_i - h.c.\right)\\
  \hat{T}_2(\bm\theta) - \hat{T}_2^{\dagger} (\bm\theta) &= \frac{1}{4} \sum_{i,j,m,n} \theta_{i,j}^{m,n} \left( \hat{a}_m^{\dagger} \hat{a}_n^{\dagger}  \hat{a}_i \hat{a}_j - h.c.\right)
\end{align}
where $i,j$ denote the occupied orbitals and $m,n$ label the unoccupied orbitals in the reference state $\vert \phi_0\rangle$. $h.c.$ is the Hermitian conjugate. Here, the excitation is allowed from occupied orbitals to unoccupied orbitals. 

In the generalized singles and doubles UCC (UCCGSD), the trial wavefunction ansatz is given by\cite{Whaley19}
\begin{equation}
    \vert \Psi(\bm\theta) \rangle= e^{\hat{T}{(\bm\theta)} - \hat{T}^{\dagger} (\bm\theta)} \vert \phi_0 \rangle 
\end{equation}
and the generalized singles and doubles excitation operators are defined as
\begin{align}
  \hat{T}_1(\bm\theta) - \hat{T}_1^{\dagger} (\bm\theta) &= \sum_{p,q} \theta_p^q \left( \hat{a}_q^{\dagger} \hat{a}_p - h.c.\right)\\
  \hat{T}_2(\bm\theta) - \hat{T}_2^{\dagger} (\bm\theta) &= \frac{1}{4} \sum_{p,q,r,s} \theta_{p,q}^{r,s} \left( \hat{a}_r^{\dagger} \hat{a}_s^{\dagger}  \hat{a}_p \hat{a}_q - h.c.\right)
\end{align}
Unlike the UCCSD, the generalized singles and doubles excitation in the UCCGSD does not distinguish between occupied and unoccupied orbitals in the reference state. Therefore, $p,q,r,s$ indices represent both types of the orbitals, occupied and unoccupied.

The $k$ unitary pair with generalized singles and doubles product wavefunction is written as\cite{Whaley19}
\begin{equation}\label{kup}
    \vert \Psi(\bm\theta) \rangle=\prod_{\delta=1}^{k} \left(e^{\hat{T}^\delta{(\bm\theta)} - \hat{T}^{\dagger \delta} (\bm\theta)}\right) \vert \phi_0 \rangle 
\end{equation}
where $k$ is an integer number that increases the wavefunction flexibility. In k-UpCCGSD, the two-body excitation is allowed to move a pair of electrons from one spatial orbital to another spatial orbital,
\begin{equation}
    \hat{T}_2(\bm\theta) - \hat{T}_2^{\dagger} (\bm\theta) = \frac{1}{4} \sum_{p,q} \theta_{p_\alpha,p_\beta}^{q_\alpha,q_\beta} \left( \hat{a}_{q_\alpha}^{\dagger} \hat{a}_{q_\alpha}^{\dagger}  \hat{a}_{p_\beta} \hat{a}_p - h.c.\right)
\end{equation}

In Table \ref{UCC-scale}, we introduce the cost of the three UCC trial wavefunctions on quantum devices using a single Trotter step. As seen, k-UpCCGSD requires the least and the UCCGSD requires the most resources. In the result and discussion section, we explain the performance of these methods to simulate gas phase models, Li$_2$Co$_2$O$_4$/Co$_2$O$_4$, which mimic the building block of the cathode battery material.

\begin{table}
\caption{\label{UCC-scale}  The cost of the UCCSD, UCCGSD, and k-UpCCGSD using a single Trotter step on quantum devices.}
\begin{tabular}{ccc}
\hline
Method & gate count\textsuperscript{\emph{a}} & circuit depth\textsuperscript{\emph{b}}\\
\hline
UCCSD& $\mathcal{O}$ $(M-N)^2$ $M^2$&$\mathcal{O}$ $(M-N)^2 M$  \\
UCCGSD& $\mathcal{O}$ ($M^4$)& $\mathcal{O}$ ($M^3$)\\
k-UpCCGSD& $\mathcal{O}$ ($kM^2$)&$\mathcal{O}$ ($kM$)\\
\hline
\end{tabular}


\textsuperscript{\emph{a}}Gate count refers to the total number of quantum gates.\\
\textsuperscript{\emph{b}}Circuit depth is the number of sequential steps to execute one- and two-qubit gates.\\
\textsuperscript{\emph{c}}$M$ denotes the number of spin-orbitals and $N$ represents the number of electrons.
\end{table}

Once the trial wavefunction is constructed, the total energy of the system can be obtained as\cite{Aspuru:16}
\begin{equation}
    E=\mathrm{min}_{\bm\theta} \,\frac{ \langle \Psi(\bm\theta) \vert \hat{H} \vert \Psi(\bm\theta) \rangle}{\langle \Psi(\bm\theta) \vert\Psi(\bm\theta) \rangle}
\end{equation}
using the variational principle. Here, $\hat{H}$ is the electronic Hamiltonian. In the second quantization formalism, the Hamiltonian in the Hartree Fock (HF) basis can be written as\cite{Szabo96,Olsen14}
\begin{equation}
    \hat {H} = \sum_{pq=1}^{N} h_{pq} \hat{a}_p^{\dagger} \hat{a}_q + \frac{1}{2} \sum_{pqrs=1} ^{N} h_{pqrs} \hat{a}_p^{\dagger} \hat{a}_q^{\dagger}  \hat{a}_r \hat{a}_s
\end{equation}
where $h_{pq}$ and $h_{pqrs}$ correspond to the one and two electron integrals, respectively
\begin{align}
    h_{pq} &= \int d\sigma\, \psi_p^{*}(\sigma) \left( - \frac{\nabla^2_{\bm{r}}}{2} - \sum_i \frac{Z_i}{\vert \bm{R}_i - \bm{r}_i \vert}\right) \, \psi_q(\sigma) \\
    h_{pqrs} & = \int \int d\sigma_1 d\sigma_2 \frac{\psi_p^{*}(\sigma_1) \psi_q^{*}(\sigma_2) \psi_r(\sigma_1) \psi_s(\sigma_2) }{\vert \bm{r}_1 - \bm{r}_2 \vert}
\end{align}
where $\bm{R}$ and $\bm{r}$ denote nuclear and electronic spatial coordinate, respectively. $\sigma$ represents the spatial and spin coordinate and $\psi(\sigma)$ is a one-electron function that is obtained from HF. 

To reduce the circuit depth and overcome the limited qubits coherence time on NISQ devices, the active space approach is employed\cite{Shepard12,Rancurel73,Olsen14}. In this approach, the Fock space of the wavefunction is divided into active (A) and inactive (I) orbitals 
\begin{equation}
    \vert \phi_0\rangle = \vert \phi_0^{A}\rangle \otimes \vert \phi_0^{I}\rangle
\end{equation}
The active orbitals include the highest occupied molecular orbitals and the lowest unoccupied molecular orbitals in a system. The assumption behind this approach is that the wavefunction is relatively dominanted by small number of determinants that one can capture by expanding the wavefunction in the active space. As a consequence, one can reduce the Hilbert space of the UCC ansatz to search to a subspace
\begin{equation}\label{u-op}
    \vert \Psi(\bm\theta) \rangle= U(\bm{\theta}) \, \vert \phi_0^{A} \rangle 
\end{equation}
where $U(\bm{\theta})$ corresponds to one of the ansatzes described above. Here, the cluster operator allows only excitations among active orbitals.  In this case, the active space Hamiltonian is given by\cite{Dyall94}
\begin{equation}\label{AH}
    \hat{H} = E_I + \sum_{tu} (t\vert\hat{f}\vert u) \, \hat{t}^{\dagger} \hat{u}+\frac{1}{2} \sum_{tuvw} (tu\vert vw)\,  \hat{t}^{\dagger}\hat{v}^{\dagger}\hat{w}\hat{u}
\end{equation}
where $t,u,v,w$ denote the active orbitals. The scalar term $E_I$ is the HF energy of the inactive orbitals. The one-electron integral is given by
\begin{equation}
    (t\vert\hat{f}\vert u) = (t\vert \hat{h} \vert u) + \sum_i 2 (tu \vert ii) - (ti \vert ui)
\end{equation}
where $i$ represents an inactive orbital. The spatial one and two-electron integrals are given in chemists' notation. 

\section{The gas-phase model and the active space}\label{AS}

\begin{figure*}
\includegraphics[width=\linewidth]{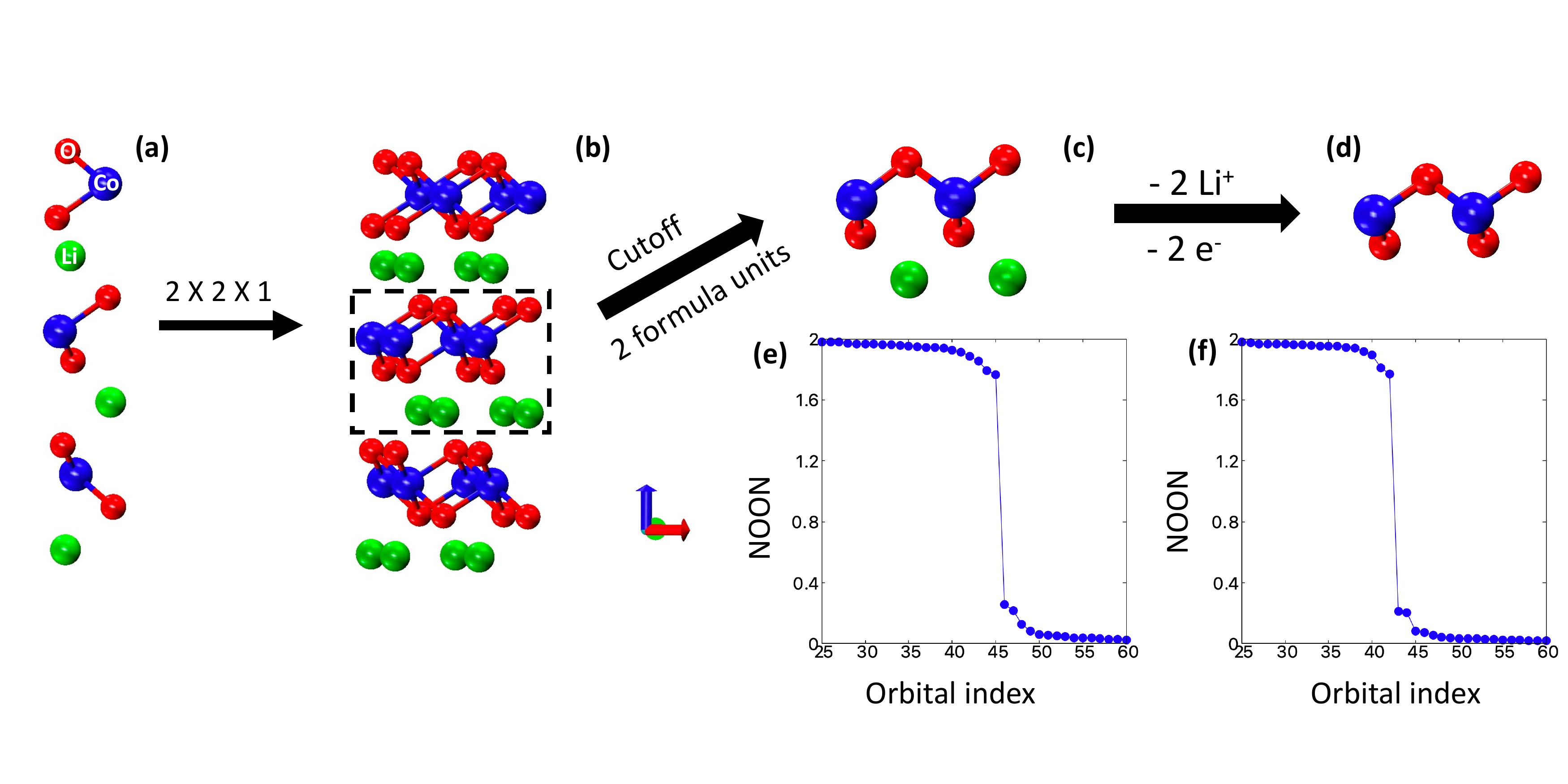}
\caption{\label{cut-off-2fu} (a) The conventional unit cell of the LiCoO$_2$. (b) The $2\times 2 \times 1$ supercell of the LiCoO$_2$. (c) The gas-phase model of the Li$_2$Co$_2$O$_4$ (d) The gas-phase model of Co$_2$O$_4$ after removing two-lithium ions and two electrons from Li$_2$Co$_2$O$_4$. (e) The natural orbital occupation numbers (NOON) as obtained from MP2/cc-pvdz for the Li$_2$Co$_2$O$_4$ and (f) Co$_2$O$_4$  gas-phase models.}
\end{figure*}

In this work, we employ gas-phase models that mimic the building blocks of the LiCoO$_2$/CoO$_2$ crystal structure. The gas-phase models are generated based on the crystal structure of LiCoO2 (R$\bar{3}$m) which we obtained from the Material Project data base\cite{Ceder04}$^,$\bibnote{https://materialsproject.org}. In Fig.\ref{cut-off-2fu}(a), we present the conventional unit cell that includes three formula units of LiCoO$_2$. Fig.\ref{cut-off-2fu}(b) shows the LiCoO$_2$ supercell comprises $2 \times 2 \times 1$ conventional unit cells and 12 formula units. We notice that every formula unit consists of one Cobalt (Co), two Oxygens (O), and one Lithium (Li). In the lithiated crystal, Co is $+3$, Li is $+1$ and O is $-2$. Therefore, the overall charge of LiCoO$_2$ is zero. In the delithiated crystal (CoO$_2$), the Li ion is removed and one electron is released. In this case, Co becomes $+4$ and the overall charge stays zero. 

To identify a small gas-phase model system for the simulation on quantum devices that mimics the LiCoO$_2$/CoO$_2$ crystal structure building block, the following criteria should be met: First, the overall charge should be maintained (for LiCoO$_2$/CoO$_2$ it is zero); second, the local oxidation state should be preserved (Co ($+3, +4$), O($-2$), and Li ($+1$)); and third, the local coordination environment for the transition metal (Co) should be conserved (every Co is coordinating with two oxygen). Evidently, these criteria represent the formula unit of LiCoO$_2$/CoO$_2$. In this work, we identify a gas-phase model which involves two formula units of the LiCoO$_2$/CoO$_2$ and they are depicted in Fig.\ref{cut-off-2fu}(c) and Fig.\ref{cut-off-2fu}(d). We should emphasize that these gas phase models do not represent the LiCoO2/CoO2 bulk material; however, they are useful starting point for estimating the performance and cost of VQE-UCC methods on transition metal oxides for battery materials. 

The constructed gas-phase model of the Li$_2$Co$_2$O$_4$ has 92 electrons and 170 molecular orbitals, while the Co$_2$O$_4$ has 86 electrons and 142 molecular orbitals using the cc-pVDZ basis set\cite{Olsen14}. Obviously, it is impossible to simulate these systems using the FCI approach on classical processors. Furthermore, we cannot explicitly simulate all electrons of these systems on quantum computers due to the limited number of qubits available in the existing quantum hardware. Hence, we decompose the Fock space into active and inactive orbitals and the active space is employed as an initial guess for the subsequent simulation on a quantum computer. This approach allows reduced circuit depth making it amenable to the currently available quantum computers.

\begin{figure}
\includegraphics[width=0.5\linewidth]{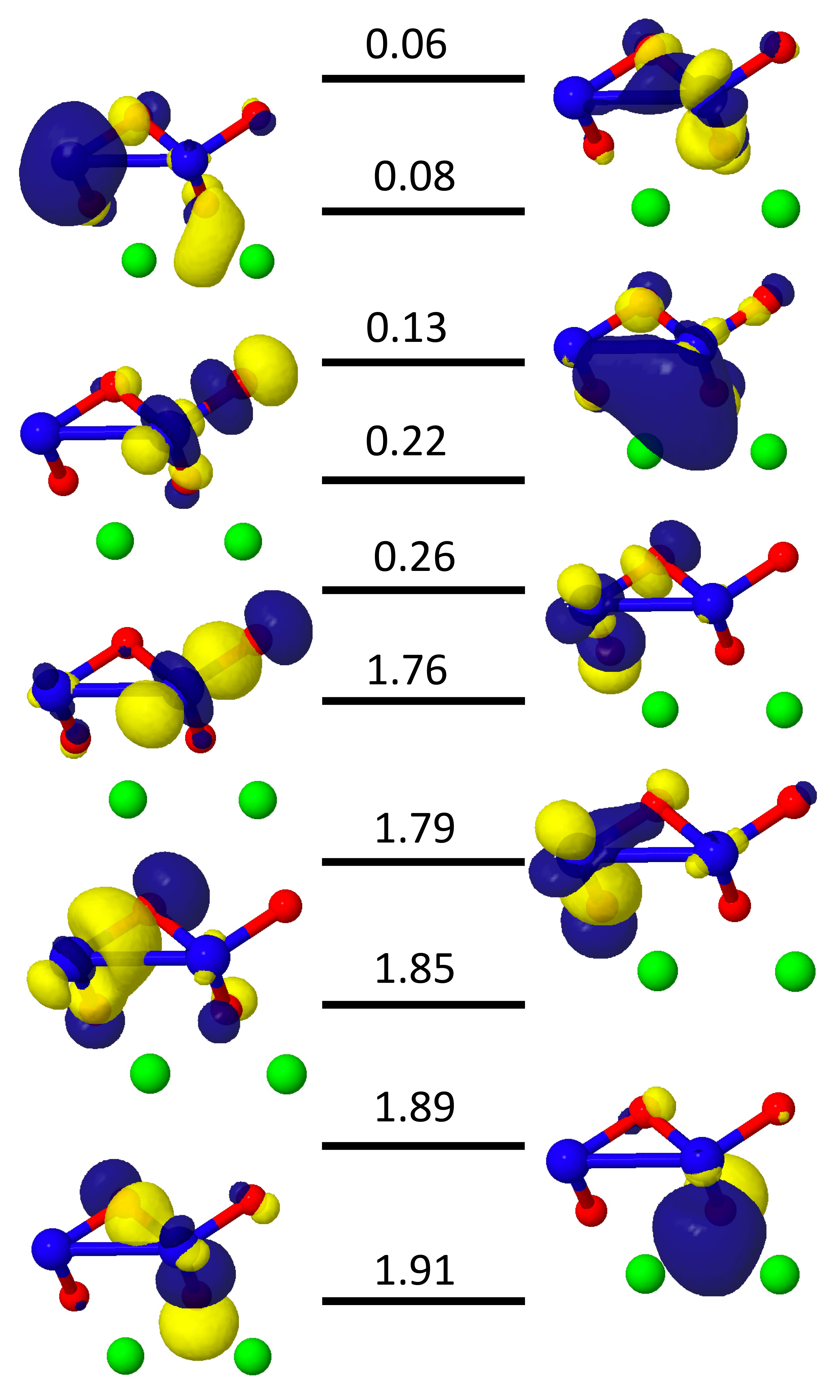}
\caption{\label{NO-LCO} The natural orbitals and the corresponding natural occupation number of the gas-phase model Li$_2$Co$_2$O$_4$ as obtained from MP2/cc-pVDZ. \textsf{Jmol}\bibnote{http://jmol.sourceforge.net} is used to plot the orbitals with cutoff=0.05.}
\end{figure}
\begin{figure}
\includegraphics[width=0.5\linewidth]{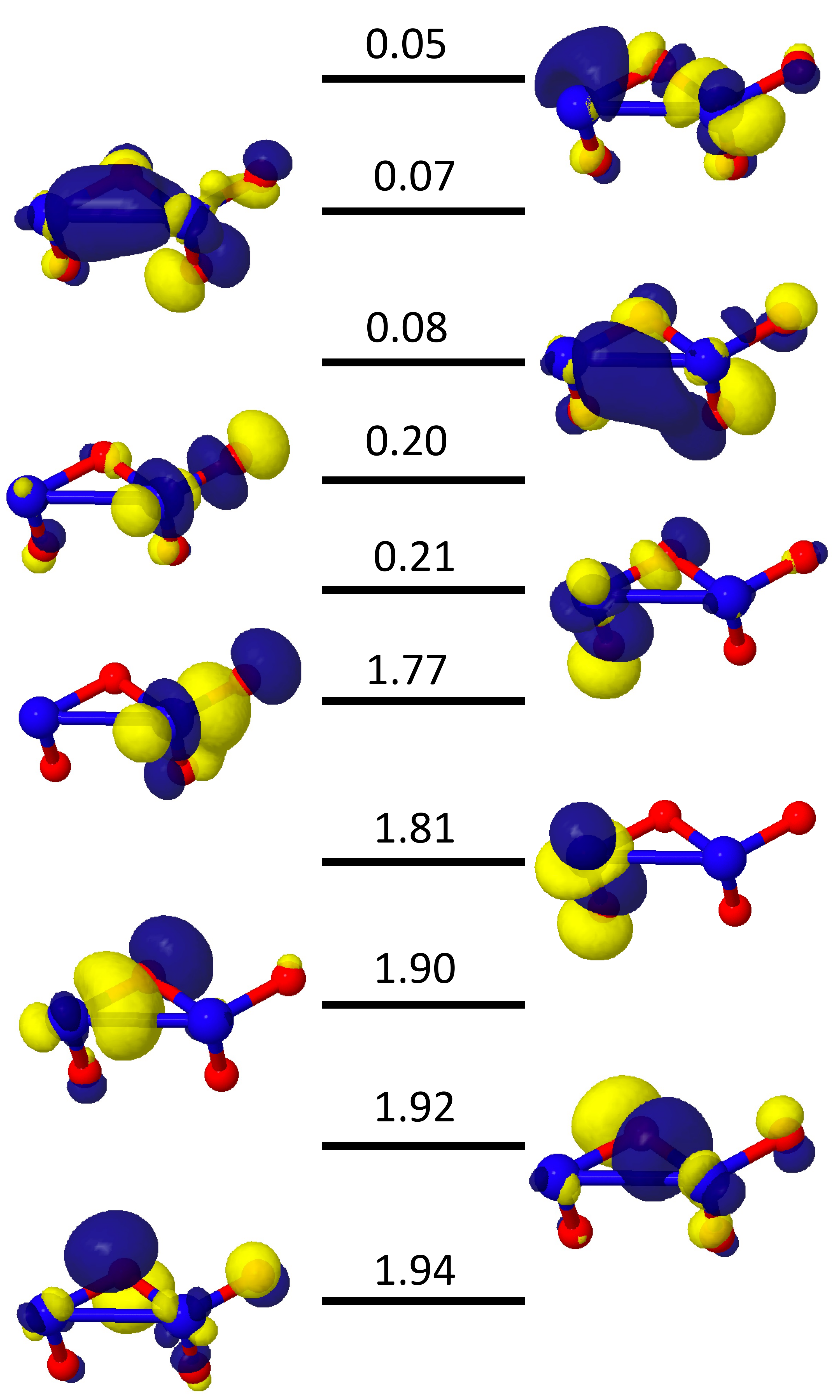}
\caption{\label{NO-CO} The natural orbitals and the corresponding natural occupation number of the gas-phase model Co$_2$O$_4$ as obtained from MP2/cc-pVDZ. \textsf{Jmol}  is used to plot the orbitals with cutoff=0.05.}
\end{figure}
To find the active space $\vert \phi_0^A\rangle$, the natural orbitals occupation numbers (NOON) are analyzed and the results are summarized in Fig.\ref{cut-off-2fu} (e) for Li$_2$Co$_2$O$_4$ and Fig.\ref{cut-off-2fu} (f) for Co$_2$O$_4$. The NOON are obtained from the diagonalization of the one-particle density matrix (1-PDM). The 1-PDM calculation is carried out using MP2/cc-pVDZ \cite{Szabo96,Olsen14,Peterson06} available in the \verb+Pyscf+ package\cite{Berkelbach20,Berkelbach18, Sun15}. MP2 accounts for the correlation of the wavefunction using perturbation theory, and therefore, it is inexpensive. Fig.\ref{cut-off-2fu} (e) and Fig.\ref{cut-off-2fu} (f) reveal that there are orbitals that are fully occupied with two-electrons, orbitals that are fully unoccupied (zero electrons), and orbitals that are semi-occupied (non-integer numbers of electrons). The orbitals with fractional occupations are considered as the active space, and the orbitals with integer occupations are considered inactive. As a consequence, the active space, in this work, comprises of 10 electrons and 10 orbitals (20 spin orbitals). These key orbitals are presented in Fig.\ref{NO-LCO} and Fig.\ref{NO-CO}. As seen, the active space is dominanted by the Co $d$-orbitals and the O $p$-orbitals. Previous experimental and theoretical works\cite{Ehrenberg09,Sawatzky91,Jaegermann10,Major18} investigating the band structure and the density of state for the crystal structure of LiCoO$_2$/CoO$_2$ reported that the valence and the lowest conduction bands are dominant by O $2p$ and Co $3d$ orbitals, and they suggested a strong hybridization between the O $2p$ and Co $3d$ orbitals. We notice that our electronic structure analyses for the gas-phase model of Li$_2$Co$_2$O$_4$ in Fig.\ref{NO-LCO} and Co$_2$O$_4$ in Fig.\ref{NO-CO} agree well with the electronic structure of the LiCoO$_2$/CoO$_2$ crystals. Therefore, we believe that the gas phase models are a good approximation for our subsequent VQE simulation. In this work, the natural orbitals of the active space are employed as an initial guess for the subsequent simulation on the quantum devices. As known, a linear transformation between natural orbitals and molecular orbitals exists via the unitary transformation matrix obtained from the diagonalization of the 1-PDM.

\begin{figure}
\includegraphics[width=0.5\linewidth]{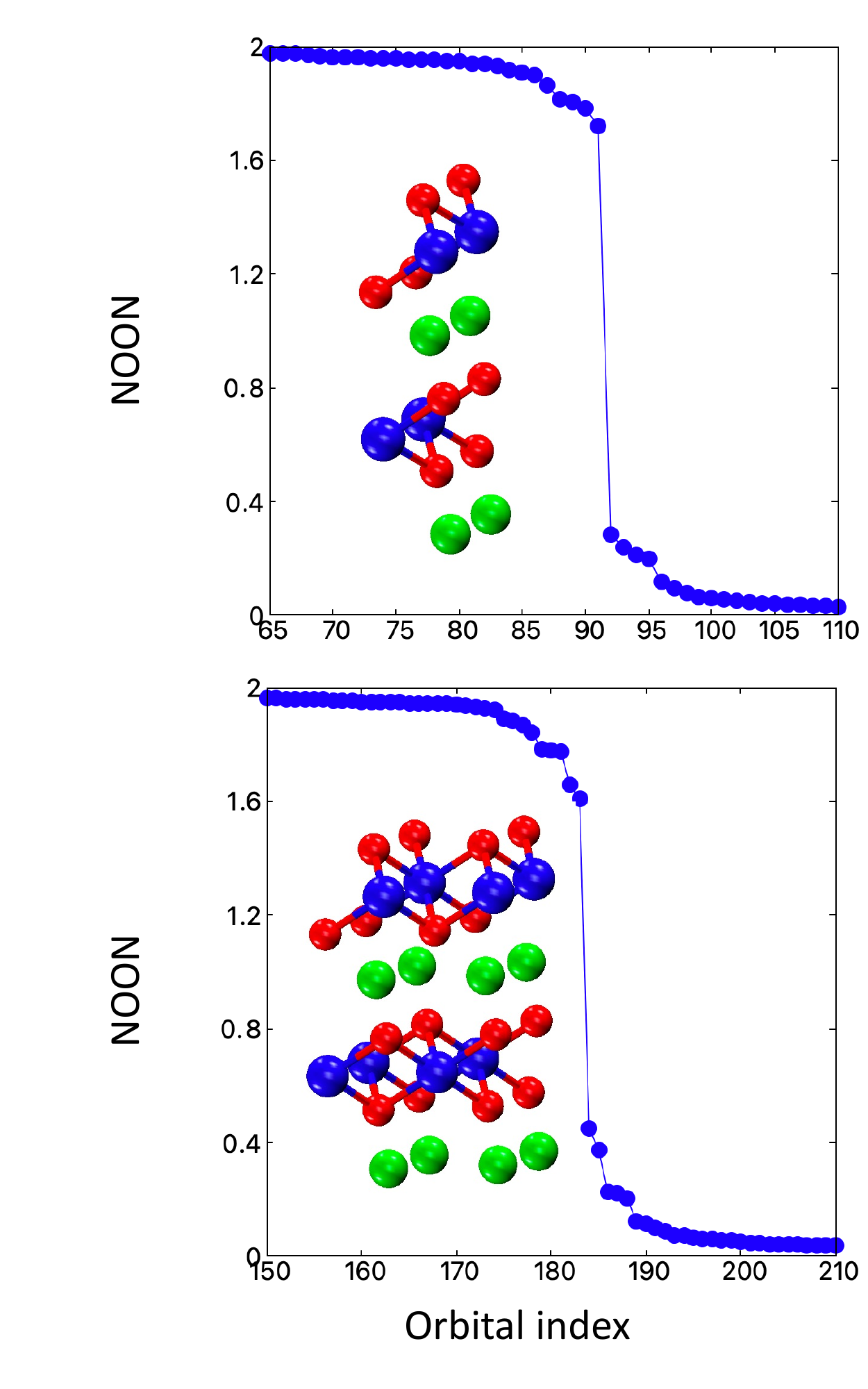}
\caption{\label{NooN-large} The natural orbital occupation numbers (NOON) as obtained from MP2/6-31g for larger gas-phase models.}
\end{figure}

Larger gas-phase models are tested as well to ensure that the results are consistent with the small gas phase models and the results are presented in Fig.\ref{NooN-large}. In these models, some of the Co ions are coordinated by six oxygen atoms. This mimics to some extent the chemical environment of Co in the bulk materials of LiCoO$_2$. We find that in these larger gas-phase models, the active space is larger, however, the active space is still dominanted by the O $p$ and Co $d$ orbitals in agreement with the small gas-phase model. Due to the resource limitations, we only considered the small gas-phase models for the VQE-UCC simulations in this work.

\section{Computational Details}\label{CD}

\begin{figure}
\includegraphics[width=0.5\linewidth]{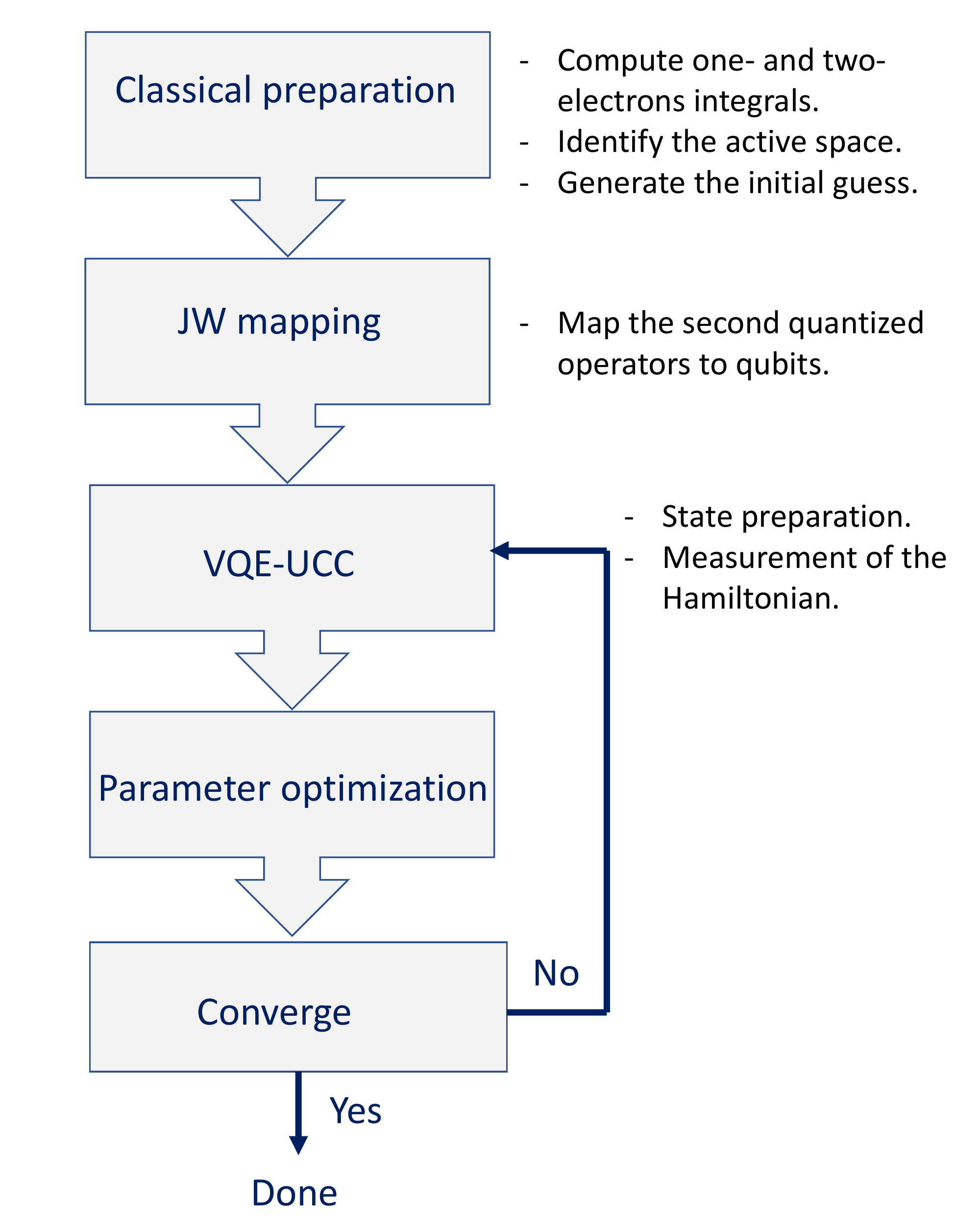}
\caption{\label{CD-scheme} Schematic representation of the variational quantum eigensolver algorithm.}
\end{figure}
Fig.\ref{CD-scheme} outlines the basic steps for our simulation using VQE algorithm. In this work, all VQE simulations are carried out using \verb+InQuanto+ package\bibnote{A. Tranter, C. Di Paola, D. M. Ramo, D. Z. Manrique, D. Gowland, G. Greene-Diniz, G. Christopoulou, I. Polyak, I. Khan, J. Pilipczuk, J. Kirsopp, K. Yamamoto, M. Tudorovskaya, M. Krompiec, N. Fitzpatrick, InQuanto - Introduction to the InQuanto Computational Chemistry Platform For Quantum Computers,  2022, https://medium.com/cambridge-quantum-computing/4fced08d66cc}. The charge and the spin ($2S+1$) for all calculations are set to zero and one, respectively. The VQE simulation steps can be summarized as follows:
\begin{itemize}
    \item Classical pre-processing. In this step, we perform all the calculations on classical processors using \verb+pyscf+ package\cite{Berkelbach20,Berkelbach18, Sun15} along with \verb+InQuanto+ interface. The cc-pVDZ is employed as the basis set for all the classical pre-processing simulations. We begin by selecting the active space (see the previous section for more details) by analyzing the natural orbitals occupation number (NOON), which obtained from the diagonalization of the 1-PDM using MP2/cc-pVDZ. Next, we calculate and store the necessary matrix elements for one- and two-electron integrals of the Hamiltonian, see Eq.(\ref{AH}), using RHF for the subsequent VQE simulations. Finally, we generate the initial guess for the vector parameter $\bm{\theta}$. Here, all initial parameters are set to zero. 
    
    \item Fermion-to-qubit mapping. Jordan-Wigner\cite{Love12,Gordon05} (JW) is employed to perform a mapping of the second-quantized operators in the Hamiltonian, see Eq.(\ref{AH}), to qubits. Here, the Hamiltonian is written as a linear combination of string of Pauli matrices. Similarly, the unitary operator ($U(\bm{\theta})$) of the UCC wavefunction (see Eq.\ref{u-op}) is transformed following the same procedure of the Hamiltonian. A single Trotter decomposition step is employed for all calculations.
    
    \item Quantum simulations on quantum devices. The trial wavefunction on the quantum device is constructed in this step. In this work, three different ansatzes based on the UCC (UCCSD, UCCGSD, k-UpCCGSD) are employed, as explained above in the theory section. Once the wavefunction is prepared, the expectation value of the Hamiltonian is computed. All calculations are carried out on the IBM statevector simulator\bibnote{https://quantum-computing.ibm.com/lab/docs/iql/manage/simulator/}  employing 20 qubits, as our active space consists of 10 electrons and 20 spin orbitals. 
    
    \item Parameter optimization. The vector parameters $\bm{\theta}$ are updated using the bounded limited-memory Broyden–Fletcher–Goldfarb–Shanno (L-BFGS-B) algorithm. An analytical gradient function of the objective function (expectation value of the Hamiltonian) for UCCSD and k-UpCCGSD is employed, while for UCCGSD a numerical gradient is employed. 
    
    \item Convergence. A stopping criterion, based on energy convergence, is selected. We export the energy once convergence is achieved. Otherwise, the updated parameters of the wavefunction $\bm{\theta}$ are used to generate a new trial wavefunction and recalculate the energy. We repeat the process until converges.
    
\end{itemize}

In addition to the VQE-UCC simulations, CASCI, CASSCF, CCSD, and MP2 calculations on classical processors are carried  out to benchmark the VQE results. All calculations are performed using \verb+pyscf+ package and cc-pVDZ basis set. The active space employed in CASCI, CASSCF, CCSD, MP2 is as same as those selected for the VQE-UCC. In CCSD, only single and double excitations are considered within the active space. Hence, CCSD span the same Hilbert space as VQE-UCC trial wavefunctions employed in this work. In CASCI and CASSCF, FCI is applied within the active space and, therefore, all configurations within the active space are treated on an equal footing. Unlike CASCI where only the CI coefficients are optimized variationally, in CASSCF both orbitals and CI coefficients are optimized variationally to give a lowest energy for the wavefunction. In MP2, a second-order perturbation theory for many-electron wavefunction is considered.

\section{Results and Discussion}\label{result}

In LIBs, the average Li intercalation potential (voltage) is calculated as\cite{Ceder10}
\begin{equation}\label{voltage}
   \langle V \rangle \simeq \frac{- \left[E(\mathrm{Li}_{x2}\mathrm{MO}_2) - E(\mathrm{Li}_{x1}\mathrm{MO}_2) - (x_2-x_1) E(\mathrm{Li}) \right] }{(x_2-x_1) e}
\end{equation}
where $E$ is the total energy and $e$ is the absolute value of the electron charge. $E(\mathrm{Li}_{x1}\mathrm{MO}_2)$ and $E(\mathrm{Li}_{x2}\mathrm{MO}_2)$ denote the total energy of the material in different oxidation states during the redox reaction. $E(\mathrm{Li})$ is the metallic (body-centered-cubic) lithium. In LiCoO$_2$ cathode battery materials, $E(\mathrm{Li}_{x1}\mathrm{MO}_2)$ can be LiCoO$_2$, where $x_1=1$, and $E(\mathrm{Li}_{x2}\mathrm{MO}_2)$ can be CoO$_2$, where $x_2=0$. This will result in voltage per formula unit. When calculating the energies using DFT, errors such as the self interaction energy error and the lack of dispersion interaction will not be canceled out, as energies here represent systems in different environments. Accordingly, DFT predicts the voltage qualitatively as reported by previous theoretical works\cite{Ceder10,Ceder16,Major20}.    

To guide the development of novel battery materials with superior performance in the absence of experimental results, accurate theoretical tools are needed that enable bridging the gap between accuracy and efficiency. Quantum computing promises to bridge this gap. In this work, we aim to understand the performance of the existing VQE-UCC approaches to simulate battery materials on the near-term quantum devices. This is a step toward establishing a universal theoretical tools to be employed as a reference value  in the absence of experimental data. 

\begin{table}
\caption{\label{E-CC-VQE} The calculated energy difference (eV) between the lithiated and delithiated gas-phase models using VQE-UCC ansatz on the IBM statevector simulator using 20 qubits. The predicted energies are benchmarked  against energies obtained from CASCI, CASSCF, CCSD, and MP2 on classical computers. }
\begin{tabular}{cccc}
\hline
Method & $E$ (eV)\textsuperscript{\emph{a}} & $dE$ (kcal/mol) & $dE$ (kcal/mol)\\
\hline
CASSCF& - 410.8311026  &-0.252199   &  0.054356      \\
CASCI & - 410.8201662  &0.0 & 0.306555\\
CCSD  & - 410.8334597  &-0.306555&0.0\\
MP2   & - 412.2219177  &-32.325149& -32.018594   \\
RHF   & - 409.9148477  &20.87713 & 21.18369\\
\hline
UCCSD  &  - 410.8122999 & 0.181401 & 0.487956\\
UCCGSD &  - 410.9984998 &-4.112470 &-3.805916\\
k-UpCCGSD (k=3)   & -410.7601677  &1.383598 &1.690153\\
k-UpCCGSD (k=5)   & -410.8672987  &-1.08690 &-0.780347\\
\hline
\end{tabular}

\textsuperscript{\emph{a}}The energy difference between the lithiated and delithiated gas-phase models  $E(\mathrm{Li}_2\mathrm{Co}_2\mathrm{O}_4)-E(\mathrm{Co}_2\mathrm{O}_4)$.
\end{table}

Table \ref{E-CC-VQE} shows the energy difference between the Li$_2$Co$_2$O$_4$ and Co$_2$O$_4$ gas phase models as obtained from UCCSD, UCCGSD, and k-UpCCGSD. The results obtained from the VQE simulation are compared against those obtained from RHF, CASCI, CASSCF, CCSD, and MP2 on classical computers. Here, the Li$_2$Co$_2$O$_4$ and Co$_2$O$_4$ gas-phase models mimic the lithiated $E(\mathrm{Li}_{x1}\mathrm{MO}_2)$ with $x_1=2$ and delithiated $E(\mathrm{Li}_{x2}\mathrm{MO}_2)$ with $x_2=0$ in the voltage calculation, see  Eq.(\ref{voltage}). In this work, we focus on the energy difference between Li$_2$Co$_2$O$_4$ and Co$_2$O$_4$ and not the voltage, because the voltage is usually calculated and measured for bulk materials instead of a simple gas-phase model. Here, the goal is to quantify the performance and predict the resources of the VQE-UCC approaches to simulate transition metal systems on quantum devices. We notice that the energy difference obtained from MP2 and UCCGSD is the lowest, while the energy difference obtained from RHF is the highest. 

To understand the results in Table \ref{E-CC-VQE}, we compute the absolute energy value for our gas phase models and the results are presented in Table \ref{abs-E}. We begin by comparing the VQE-UCC results with the CCSD and CASCI results. In case of Li$_2$Co$_2$O$_4$, we find that the ground state (GS) energy obtained from RHF is higher than those obtained from CASCI and CCSD by 186.45 kcal/mol and 182.51 kcal/mol, respectively. This is expected because in RHF, the wavefunction is described as a one single determinant\cite{Szabo96}. Consequently, the dynamic and static correlations are missing. On the other hand, the GS energy obtained from the VQE-UCC ansatz agrees better with CCSD than those obtained from CASCI. As seen, the GS energy obtained from UCCSD differs from those obtained from CASCI and CCSD by 3.97 kcal/mol and 0.02 kcal/mol, respectively. Unlike UCCSD, k-UpCCGSD (k=3) results exihibt less agreement with the results obtained from CCSD and CASCI. We find that the GS energy obtained from k-UpCCGSD (k=3) is higher than the GS energy obtained from CASCI and CCSD by 7.17 kcal/mol and 3.22 kcal/mol, respectively. However, adding more k product (k-UpCCGSD (k=5)) brings the results of k-UpCCGSD closer to CCSD and CASCI, because it enhances the wavefunction flexibility (see Eq. (\ref{kup})). We find that the GS energy obtained from k-UpCCGSD (k=5) differs from those obtained using CASCI and CCSD by 3.46 kcal/mol and -0.48 kcal/mol, respectively. Surprisingly, we notice that the GS energy obtained from UCCGSD is higher than those obtained from CASCI and CCSD by 5.56 kcal/mol and 1.62 kcal/mol, respectively. This is unexpected as UCCGSD wavefunction is more flexible than UCCSD. One reason for this could be due to the employment of a numerical gradient to calculate the objective function (expectation value of the Hamiltonian) for the optimization process during the VQE simulation because of the resource limitations. The numerical gradient is likely increasing the numerical error. In contrast to UCCGSD, in UCCSD and k-UpCCGSD, we employ analytical solution to calculate the objective function. Table \ref{abs-E} also introduces the GS energy values of the Co$_2$O$_4$. As we can see, similar results to Li$_2$Co$_2$O$_4$ are observed in Co$_2$O$_4$.

\begin{table*}
\caption{\label{abs-E} The calculated ground state energy in hartree for Li$_2$Co$_2$O$_4$ and Co$_2$O$_4$ using VQE-UCC on IBM statevector simulator and RHF, CCSD, MP2, CASCI, and CASSCF on classical processors.}
\scalebox{0.8}{
\begin{tabular}{ccccccc}
\hline
 Method&E(Li$_2$Co$_2$O$_4$)&dE(kcal/mol)&dE(kcal/mol)&E(Co$_2$O$_4$ )&dE(kcal/mol)&dE(kcal/mol)\\
 \hline
 CASSCF&-3077.021047220& -112.146059 & -116.089857&-3061.923289747&-111.893859& -116.144213\\
 CASCI &-3076.842331137&0.0&-3.943799&-3061.744975570&0.0&-4.250354\\
 CCSD&-3076.836046295&3.94379928&0.0&-3061.738202201&4.250354&0.0\\
 MP2 &-3076.920237209& -48.886809 &-52.830607&-3061.771368249&-16.5616596& -20.8120138 \\
 RHF&-3076.545202892&186.450829&182.507030&-3061.481117150&165.573694&161.323340\\
 \hline
 UCCSD&-3076.836011187& 3.965830&0.022030&-3061.738944700&3.784429&-0.465925\\
 UCCGSD&-3076.833469294& 5.560892&1.617092&-3061.729560089&9.673362&5.423008\\
 k-UpCCGSD (k=3)&-3076.830912759&7.165142&3.221343&-3061.735762095&5.781544&1.531190\\
 k-UpCCGSD (k=5)&-3076.836817725&3.459719&-0.484080&-3061.737730070&4.546620&0.296267\\
\hline
\end{tabular}}
\end{table*}

Transition metal oxides, such as the simple gas-phase model employed in this paper, are known to be strongly correlated systems\cite{Rubio12,Sadovskii13,Pacchioni15,Temmerman08,Friesner19}. In our results, we observe that at equilibrium geometry, the results obtained from CCSD are slightly higher than those obtained from the CASCI by 3.94 kcal/mol and 4.25 kcal/mol in Li$_2$Co$_2$O$_4$ and Co$_2$O$_4$, respectively. In VQE-UCC, we notice that the energy difference between UCCSD and CASCI is 3.97 kcal/mol and 3.78 kcal/mol in Li$_2$Co$_2$O$_4$ and Co$_2$O$_4$, respectively. Similar values are observed for k-UpCCGSD (k=5). To further understand these results, we compare them against those obtained from MP2 and CASSCF. We note that the CASCI and CCSD GS energies are higher than CASSCF GS energy by 112.15 kcal/mol and 116.09 kcal/mol, repectively. In addition, the CASCI and CCSD GS energies excess MP2 GS energies by  48.89 kcal/mol and 52.83 kcal/mol, respectively. The qualitative agreement between VQE-UCC and CASCI, CASSCF, and MP2 could be due to two reasons. First, the missing correlation from three and more electron excitations in the cluster operator\cite{Oliveira17,Truhlar15,Gordon19}; and second, the multireference character of the wavefunction\cite{Musial07,Gordon19,Oliveira17}. These speculations are valuable starting points for further studies in the future.

Now, we analyze the energy difference between the lithiated and delithiated gas-phase models, see Table \ref{E-CC-VQE}. Here, we expect that error cancellation will improve the agreement between VQE-UCC method and CASSCF and CASCI. Indeed, in Table \ref{E-CC-VQE}, we notice that except for UCCGSD, the values obtained from CCSD and VQE-UCC are quantitatively agree with those obtained from CASCI and CASSCF. For instance, the energy obtained from CCSD is slightly lower than that obtained from CASCI by 0.31 kcal/mol and CASSCF by 0.05 kcal/mol, and in UCCSD, the energy obtained is slightly above that obtained from the CASCI by 0.18 kcal/mol and CASSCF by 0.43 kcal/mol. Similarly, in k-UpCCGSD (k=5), the values are slightly lower than that obtained from  CASCI by 1.09 kcal/mol and CASSCF by 0.83 kcal/mol.  In Table\ref{E-CC-VQE}, the error cancellation is responsible for the better performance of the CCSD and VQE-UCC methods with respect to CASCI and CASSCF.   

\begin{table*}
\caption{\label{Res-Est} resource estimation to simulate the gas-phase model Li2Co2O4 and Co2O4 on the IBM statevector simulator using 20 qubits.}
\begin{tabular}{ccccc}
\hline
 Method & Circuit depth\textsuperscript{\emph{a}} & Circuit gates & Cnot gates & Parameters\\
 \hline
 UCCSD &32806 & 99910& 20950&675 \\
 UCCGSD& 118604& 344470& 78150&2955 \\
 k-UpCCGSD (k=3)& 7113& 43390&5850 &405 \\
 k-UpCCGSD (k=5)& 11765& 72310&9750 &675 \\
 \hline
\end{tabular}

\textsuperscript{\emph{a}} Number of steps required to execute one-qubit and two-qubit gates on the circuit.
\end{table*}

Next, we investigate the cost and resources needed to simulate Li$_2$Co$_2$O$_4$ and Co$_2$O$_4$ gas-phase models using VQE-UCC approaches under the assumption of single reference state on quantum devices. Table \ref{Res-Est} presents the resources in terms of circuit depth, circuit gates, cnot gates and the number of the wavefunction parameters $\theta$. Circuit depth and gates are important metrics for the NISQ devices due to the limited coherence time of existing qubits. Our analyses using 20 qubits show that k-UpCCGSD has the lowest circuit depth and number of gates compare to the UCCSD and UCCGSD. Results in Table \ref{E-CC-VQE} and Table \ref{Res-Est} indicate that   k-UpCCGSD (k=5) produces results that are similar to UCCSD but at lower cost. As seen, the circuit depth for the k-UpCCGSD are 11765 compared to 32806 in UCCSD. Furthermore, the number of gates needed in k-UpCCGSD (k=5) wavefunction ansatz is 72310 compared to 99910 in UCCSD. For the wavefunction parameters, we notice that both UCCSD and k-UpCCGSD (k=5) require the same number of parameters. 

\begin{figure}
\includegraphics[width=0.5\linewidth]{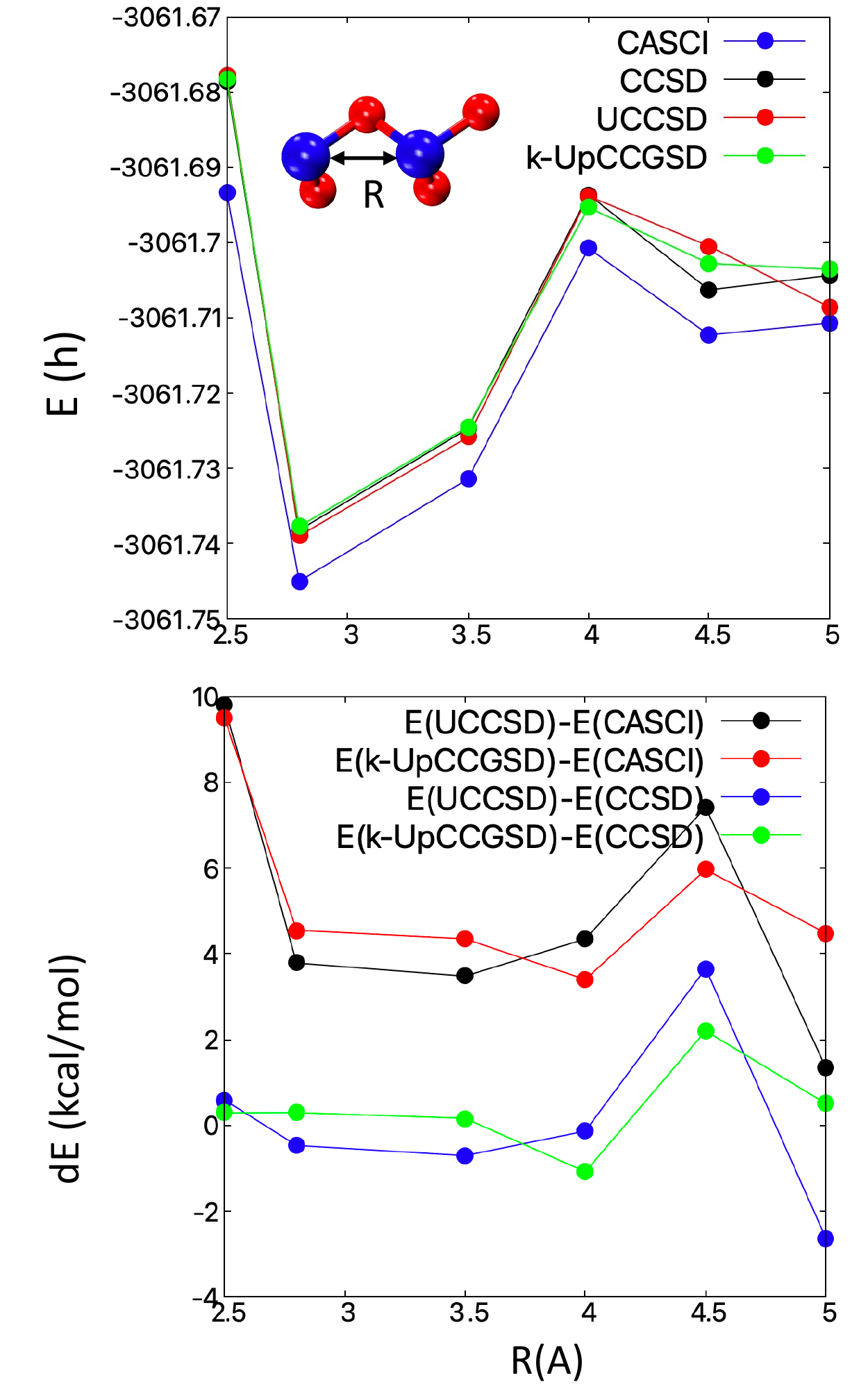}
\caption{\label{E-dis} Upper panel: The calculated energy of the Co$_2$O$_4$ using the UCCSD and k-UpCCGSD (k=5) using VQE method and the conventional CCSD and CASCI on the classical computer. Lower panel: The difference between the calculated energy using UCCSD, k-UpCCGSD(k=5) and CCSD.}
\end{figure}

Finally, we analyze the energy as a function of the distance between the two formula units. In Fig.\ref{E-dis}, the potential energy curve of the Co$_2$O$_4$ as a function of the distance between the two formula units is presented. The equilibrium distance, here, is 2.8\AA. We compare the results obtained from the UCCSD and k-UpCCGSD (k=5) with those obtained from CASCI and CCSD. Fig.\ref{E-dis} uncovers that the potential energy curve obtained from CCSD and VQE-UCC qualitatively agree with that obtained from CASCI. This is likely due to the two reasons explained above. When comparing the VQE-UCC and CCSD results, we find that at short distance (2.5-4.0 \AA) between the two formula units, the results obtained from VQE-UCC vary almost by 0.5 kcal/mol from that obtained using CCSD.  However, at longer distance (4.5-5.0 \AA) the results vary significantly ($>1$ kcal/mol) with respect to CCSD.  One reason for this could be the open-shell character of the Co$^{+4}$, when the two Co ions become far from each other. In Co$_2$O$_4$, the Co ion is ${+4}$ and the electronic configuration is [Ar] $3d^5$ ($t_2g$: 5 and $e_g$: 0). At longer distance, a multi-reference state as an initial guess is likely needed. This will be a study for future work.

Overall, our results reveal that the VQE-UCC (UCCSD and k-UpCCGSD) approaches cannot reproduce quantitatively the CASCI and CASSCF GS energy for Li$_2$Co$_2$O$_4$ and Co$_2$O$_4$ gas-phase models and this is likely because UCCSD and k-UpCCGSD do not capture all the correlation in the total wavefunctions.  Employing  multireference wavfunctions as initial guesses and extension of VQE-UCC to three and more electron excitations are likely to improve the results of VQE-UCC approaches.  In quantum computers we can take advantage of the quantum nature of the hardware (superposition, entanglement, and interference) to advance the VQE-UCC and implement the missing correlation and at the same time keep the VQE-UCC at reasonable scaling. Moreover, our results suggest that we need 20 qubits to simulate two formula units of LiCoO2 and CoO2. Replicating these two formula units along the $x$, $y$, and $z$ direction to generate a supercell of 40 formula units might require approximately 400 qubits that is likely to be available in the near future. Therefore, VQE method is likely to be a crucial tool to simulate strongly correlated systems of large size and provide insights as the quantum hardware matures.  


\section {Conclusions}\label{summary}

We simulated Li$_2$Co$_2$O$_4$ and Co$_2$O$_4$ gas-phase models to estimate the performance of the VQE-UCC (UCCSD, UCCGSD, kUpCCGSD) approaches and compute the resources needed to simulate cathode battery materials on near-term quantum devices. Unlike H$_2$, LiH, N$_2$, and H$_2$O that have been employed as  benchmark systems to assess the performance of the VQE method in earlier works, we employed the transition metal-oxide systems, Li$_2$Co$_2$O$_4$ and Co$_2$O$_4$, which are known to be strongly correlated systems. In this study, we compared the GS energy obtained from the VQE-UCC (UCCSD, UCCGSD, kUpCCGSD) approaches with those obtained from classical wavefunction-based approaches, such as CASCI, CASSCF, CCSD, and MP2. We found that while the GS energy obtained from VQE-UCC quantitatively agree with those obtained from CCSD, they only qualitatively agree with those obtained from CASCI, CASSCF, and MP2. The reasons could be due to the missing correlation from high-level electronic excitations in the cluster operator and the multireference character of the wavefunction. Since the single reference state, such as HF state, has no sufficient overlap with the true ground state for these systems, an entangled reference state that captures the states contributing to the total wavefunction might be necessary. Therefore, more advanced trial wavefunctions need to be developed to simulate such strongly correlated systems. Future works will focus on developing methods that allow simulating transition metal systems and capture the missing correlations.

\begin{acknowledgement}
The authors thank Devesh Upadhyay and Andrew Getsoian for fruitful discussion and for reviewing the manuscript.
\end{acknowledgement}




\bibliography{achemso-demo}

\providecommand{\latin}[1]{#1}
\makeatletter
\providecommand{\doi}
  {\begingroup\let\do\@makeother\dospecials
  \catcode`\{=1 \catcode`\}=2 \doi@aux}
\providecommand{\doi@aux}[1]{\endgroup\texttt{#1}}
\makeatother
\providecommand*\mcitethebibliography{\thebibliography}
\csname @ifundefined\endcsname{endmcitethebibliography}
  {\let\endmcitethebibliography\endthebibliography}{}
\begin{mcitethebibliography}{99}
\providecommand*\natexlab[1]{#1}
\providecommand*\mciteSetBstSublistMode[1]{}
\providecommand*\mciteSetBstMaxWidthForm[2]{}
\providecommand*\mciteBstWouldAddEndPuncttrue
  {\def\EndOfBibitem{\unskip.}}
\providecommand*\mciteBstWouldAddEndPunctfalse
  {\let\EndOfBibitem\relax}
\providecommand*\mciteSetBstMidEndSepPunct[3]{}
\providecommand*\mciteSetBstSublistLabelBeginEnd[3]{}
\providecommand*\EndOfBibitem{}
\mciteSetBstSublistMode{f}
\mciteSetBstMaxWidthForm{subitem}{(\alph{mcitesubitemcount})}
\mciteSetBstSublistLabelBeginEnd
  {\mcitemaxwidthsubitemform\space}
  {\relax}
  {\relax}

\bibitem[Etacheri \latin{et~al.}(2011)Etacheri, Marom, Elazari, Salitra, and
  Aurbach]{Aurbach11}
Etacheri,~V.; Marom,~R.; Elazari,~R.; Salitra,~G.; Aurbach,~D. \emph{Energy \
  Environ.\ Sci.} \textbf{2011}, \emph{4}, 3243\relax
\mciteBstWouldAddEndPuncttrue
\mciteSetBstMidEndSepPunct{\mcitedefaultmidpunct}
{\mcitedefaultendpunct}{\mcitedefaultseppunct}\relax
\EndOfBibitem
\bibitem[Thackeray \latin{et~al.}(2012)Thackeray, Wolverton, and
  Isaacs]{Isaacs12}
Thackeray,~M.~M.; Wolverton,~C.; Isaacs,~E.~D. \emph{Energy \ Environ.\ Sci.}
  \textbf{2012}, \emph{5}, 7854\relax
\mciteBstWouldAddEndPuncttrue
\mciteSetBstMidEndSepPunct{\mcitedefaultmidpunct}
{\mcitedefaultendpunct}{\mcitedefaultseppunct}\relax
\EndOfBibitem
\bibitem[Winter \latin{et~al.}(2018)Winter, Barnett, and Xu]{Xu14}
Winter,~M.; Barnett,~B.; Xu,~K. \emph{Chem.\ Rev.} \textbf{2018}, \emph{118},
  11433\relax
\mciteBstWouldAddEndPuncttrue
\mciteSetBstMidEndSepPunct{\mcitedefaultmidpunct}
{\mcitedefaultendpunct}{\mcitedefaultseppunct}\relax
\EndOfBibitem
\bibitem[Chen \latin{et~al.}(2020)Chen, Li, Yu, Chen, and Li]{Li20}
Chen,~R.; Li,~Q.; Yu,~X.; Chen,~L.; Li,~H. \emph{Chem.\ Rev.} \textbf{2020},
  \emph{120}, 6820\relax
\mciteBstWouldAddEndPuncttrue
\mciteSetBstMidEndSepPunct{\mcitedefaultmidpunct}
{\mcitedefaultendpunct}{\mcitedefaultseppunct}\relax
\EndOfBibitem
\bibitem[Tomaszewska \latin{et~al.}(2019)Tomaszewska, Chu, Feng, O'Kane, Liu,
  Chen, Ji, Endler, Li, Liu, Li, Zheng, Vetterlein, Gao, Du, Parkes, Ouyang,
  Marinescu, Offer, and Wu]{Wu19}
Tomaszewska,~A. \latin{et~al.}  \emph{eTransportation} \textbf{2019}, \emph{1},
  100011\relax
\mciteBstWouldAddEndPuncttrue
\mciteSetBstMidEndSepPunct{\mcitedefaultmidpunct}
{\mcitedefaultendpunct}{\mcitedefaultseppunct}\relax
\EndOfBibitem
\bibitem[Winter and Brodd(2004)Winter, and Brodd]{Winter04}
Winter,~M.; Brodd,~R.~J. \emph{Chem.\ Rev.} \textbf{2004}, \emph{104},
  4245\relax
\mciteBstWouldAddEndPuncttrue
\mciteSetBstMidEndSepPunct{\mcitedefaultmidpunct}
{\mcitedefaultendpunct}{\mcitedefaultseppunct}\relax
\EndOfBibitem
\bibitem[Manthiram(2017)]{Manthiram17}
Manthiram,~A. \emph{Cent.\ Sci.} \textbf{2017}, \emph{3}, 1063\relax
\mciteBstWouldAddEndPuncttrue
\mciteSetBstMidEndSepPunct{\mcitedefaultmidpunct}
{\mcitedefaultendpunct}{\mcitedefaultseppunct}\relax
\EndOfBibitem
\bibitem[Whittingham(2004)]{Whittingham04}
Whittingham,~M.~S. \emph{Chem.\ Rev.} \textbf{2004}, \emph{104}, 4271\relax
\mciteBstWouldAddEndPuncttrue
\mciteSetBstMidEndSepPunct{\mcitedefaultmidpunct}
{\mcitedefaultendpunct}{\mcitedefaultseppunct}\relax
\EndOfBibitem
\bibitem[Googenough and Park(2013)Googenough, and Park]{Park13}
Googenough,~J.~B.; Park,~K.-S. \emph{J.\ Am.\ Chem.\ Soc.} \textbf{2013},
  \emph{135}, 1167\relax
\mciteBstWouldAddEndPuncttrue
\mciteSetBstMidEndSepPunct{\mcitedefaultmidpunct}
{\mcitedefaultendpunct}{\mcitedefaultseppunct}\relax
\EndOfBibitem
\bibitem[Tian \latin{et~al.}(2021)Tian, Zeng, Rutt, Shi, Kim, Wang, Koettgen,
  Sun, Ouyang, Chen, Lun, Rong, and Persson]{Ceder21}
Tian,~Y.; Zeng,~G.; Rutt,~A.; Shi,~T.; Kim,~H.; Wang,~J.; Koettgen,~J.;
  Sun,~Y.; Ouyang,~B.; Chen,~T.; Lun,~Z.; Rong,~Z.; Persson,~K. \emph{Chem.\
  Rev.} \textbf{2021}, \emph{121}, 1623\relax
\mciteBstWouldAddEndPuncttrue
\mciteSetBstMidEndSepPunct{\mcitedefaultmidpunct}
{\mcitedefaultendpunct}{\mcitedefaultseppunct}\relax
\EndOfBibitem
\bibitem[Nitta \latin{et~al.}(2015)Nitta, Wu, Lee, and Yushin]{Yushin15}
Nitta,~N.; Wu,~F.; Lee,~J.~T.; Yushin,~G. \emph{Materials Today} \textbf{2015},
  \emph{18}, 252\relax
\mciteBstWouldAddEndPuncttrue
\mciteSetBstMidEndSepPunct{\mcitedefaultmidpunct}
{\mcitedefaultendpunct}{\mcitedefaultseppunct}\relax
\EndOfBibitem
\bibitem[Mishra \latin{et~al.}(2018)Mishra, Mehta, Basu, Malode, Shetti,
  Shukla, Nadagouda, and Aminabhavi]{Aminabhavi18}
Mishra,~A.; Mehta,~A.; Basu,~S.; Malode,~S.~J.; Shetti,~N.~P.; Shukla,~S.~S.;
  Nadagouda,~M.~N.; Aminabhavi,~T.~M. \emph{Materials Science for Energy
  Technologies} \textbf{2018}, \emph{1}, 182\relax
\mciteBstWouldAddEndPuncttrue
\mciteSetBstMidEndSepPunct{\mcitedefaultmidpunct}
{\mcitedefaultendpunct}{\mcitedefaultseppunct}\relax
\EndOfBibitem
\bibitem[Wang \latin{et~al.}(2020)Wang, Fu, Kammampata, McOwen, Samson, Zhang,
  Hitz, Nolan, Wachsman, Mo, Thangadurai, and Hu]{Hu20}
Wang,~C.; Fu,~K.; Kammampata,~S.~P.; McOwen,~D.~W.; Samson,~A.~J.; Zhang,~L.;
  Hitz,~G.~T.; Nolan,~A.~M.; Wachsman,~E.~D.; Mo,~Y.; Thangadurai,~V.; Hu,~L.
  \emph{Chem.\ Rev.} \textbf{2020}, \emph{120}, 4257\relax
\mciteBstWouldAddEndPuncttrue
\mciteSetBstMidEndSepPunct{\mcitedefaultmidpunct}
{\mcitedefaultendpunct}{\mcitedefaultseppunct}\relax
\EndOfBibitem
\bibitem[Li \latin{et~al.}(2020)Li, Wang, Chen, Xu, and Lu]{Lu20}
Li,~M.; Wang,~C.; Chen,~Z.; Xu,~K.; Lu,~J. \emph{Chem.\ Rev.} \textbf{2020},
  \emph{120}, 6783\relax
\mciteBstWouldAddEndPuncttrue
\mciteSetBstMidEndSepPunct{\mcitedefaultmidpunct}
{\mcitedefaultendpunct}{\mcitedefaultseppunct}\relax
\EndOfBibitem
\bibitem[Fleischmann \latin{et~al.}(2020)Fleischmann, Mitchell, Wang, Zhan,
  en~Jiang, Presser, and Augustyn]{Augustyn20}
Fleischmann,~S.; Mitchell,~J.~B.; Wang,~R.; Zhan,~C.; en~Jiang,~D.;
  Presser,~V.; Augustyn,~V. \emph{Chem.\ Rev.} \textbf{2020}, \emph{120},
  6738\relax
\mciteBstWouldAddEndPuncttrue
\mciteSetBstMidEndSepPunct{\mcitedefaultmidpunct}
{\mcitedefaultendpunct}{\mcitedefaultseppunct}\relax
\EndOfBibitem
\bibitem[Xu \latin{et~al.}(2017)Xu, Lin, Doeff, and Tong]{Tong17}
Xu,~J.; Lin,~F.; Doeff,~M.~M.; Tong,~W. \emph{J.\ Mater.\ Chem.\ A}
  \textbf{2017}, \emph{5}, 874\relax
\mciteBstWouldAddEndPuncttrue
\mciteSetBstMidEndSepPunct{\mcitedefaultmidpunct}
{\mcitedefaultendpunct}{\mcitedefaultseppunct}\relax
\EndOfBibitem
\bibitem[Kim \latin{et~al.}(2018)Kim, Jun, Park, Zhang, Kaghazchi, Aurbach,
  Major, Goobes, Dixit, Leifer, Wang, Yan, Ahn, Kim, Yoon, and Sun]{Sun18}
Kim,~U.-H. \latin{et~al.}  \emph{Energy\ Environ.\ Sci.} \textbf{2018},
  \emph{11}, 1271\relax
\mciteBstWouldAddEndPuncttrue
\mciteSetBstMidEndSepPunct{\mcitedefaultmidpunct}
{\mcitedefaultendpunct}{\mcitedefaultseppunct}\relax
\EndOfBibitem
\bibitem[Bak \latin{et~al.}(2014)Bak, Hu, Zhou, Yu, Sennanayake, Cho, Kim,
  Chung, Yang, and Nam]{Nam14}
Bak,~S.-M.; Hu,~E.; Zhou,~Y.; Yu,~X.; Sennanayake,~S.~D.; Cho,~S.-J.;
  Kim,~K.-B.; Chung,~K.~Y.; Yang,~X.-Q.; Nam,~K.-W. \emph{App.\ Mater.\
  Interface} \textbf{2014}, \emph{6}, 22594\relax
\mciteBstWouldAddEndPuncttrue
\mciteSetBstMidEndSepPunct{\mcitedefaultmidpunct}
{\mcitedefaultendpunct}{\mcitedefaultseppunct}\relax
\EndOfBibitem
\bibitem[Zhou \latin{et~al.}(2022)Zhou, Chen, and Cui]{Cui22}
Zhou,~G.; Chen,~H.; Cui,~Y. \emph{Nature\ Energy} \textbf{2022}, \emph{7},
  312\relax
\mciteBstWouldAddEndPuncttrue
\mciteSetBstMidEndSepPunct{\mcitedefaultmidpunct}
{\mcitedefaultendpunct}{\mcitedefaultseppunct}\relax
\EndOfBibitem
\bibitem[Chakraborty \latin{et~al.}(2020)Chakraborty, Kunnikuruvan, Kumar,
  Markovsky, Aurbach, Dixit, and Major]{Major20}
Chakraborty,~A.; Kunnikuruvan,~S.; Kumar,~S.; Markovsky,~B.; Aurbach,~D.;
  Dixit,~M.; Major,~D.~T. \emph{Chem.\ Mater} \textbf{2020}, \emph{32},
  915\relax
\mciteBstWouldAddEndPuncttrue
\mciteSetBstMidEndSepPunct{\mcitedefaultmidpunct}
{\mcitedefaultendpunct}{\mcitedefaultseppunct}\relax
\EndOfBibitem
\bibitem[der Ven \latin{et~al.}(2020)der Ven, Deng, Banerjee, and Ong]{Ong20}
der Ven,~A.~V.; Deng,~Z.; Banerjee,~S.; Ong,~S.~P. \emph{chem.\ Rev.}
  \textbf{2020}, \emph{120}, 6977\relax
\mciteBstWouldAddEndPuncttrue
\mciteSetBstMidEndSepPunct{\mcitedefaultmidpunct}
{\mcitedefaultendpunct}{\mcitedefaultseppunct}\relax
\EndOfBibitem
\bibitem[Islam and Fisher(2014)Islam, and Fisher]{Fisher14}
Islam,~M.~S.; Fisher,~C. A.~J. \emph{Chem.\ Soc.\ Rev.} \textbf{2014},
  \emph{43}, 185\relax
\mciteBstWouldAddEndPuncttrue
\mciteSetBstMidEndSepPunct{\mcitedefaultmidpunct}
{\mcitedefaultendpunct}{\mcitedefaultseppunct}\relax
\EndOfBibitem
\bibitem[Loftager \latin{et~al.}(2016)Loftager, Gracia-Lastra, and
  Vegge]{Vegg16}
Loftager,~S.; Gracia-Lastra,~J.~M.; Vegge,~T. \emph{J. Phys. Chem. C}
  \textbf{2016}, \emph{120}, 18355\relax
\mciteBstWouldAddEndPuncttrue
\mciteSetBstMidEndSepPunct{\mcitedefaultmidpunct}
{\mcitedefaultendpunct}{\mcitedefaultseppunct}\relax
\EndOfBibitem
\bibitem[Meng and Dompablo(2013)Meng, and Dompablo]{DOMPABLO13}
Meng,~Y.~S.; Dompablo,~M. E. A.-D. \emph{Acc.\ Chem. Res.} \textbf{2013},
  \emph{46}, 1171\relax
\mciteBstWouldAddEndPuncttrue
\mciteSetBstMidEndSepPunct{\mcitedefaultmidpunct}
{\mcitedefaultendpunct}{\mcitedefaultseppunct}\relax
\EndOfBibitem
\bibitem[Urban \latin{et~al.}(2016)Urban, Seo, and Ceder]{Ceder16}
Urban,~A.; Seo,~D.-H.; Ceder,~G. \emph{npj Computational Materials}
  \textbf{2016}, 16002\relax
\mciteBstWouldAddEndPuncttrue
\mciteSetBstMidEndSepPunct{\mcitedefaultmidpunct}
{\mcitedefaultendpunct}{\mcitedefaultseppunct}\relax
\EndOfBibitem
\bibitem[Lacivita \latin{et~al.}(2018)Lacivita, Artrith, and Ceder]{Ceder18}
Lacivita,~V.; Artrith,~N.; Ceder,~G. \emph{Chem. Mater.} \textbf{2018},
  \emph{30}, 7077\relax
\mciteBstWouldAddEndPuncttrue
\mciteSetBstMidEndSepPunct{\mcitedefaultmidpunct}
{\mcitedefaultendpunct}{\mcitedefaultseppunct}\relax
\EndOfBibitem
\bibitem[Whitfield \latin{et~al.}(2013)Whitfield, Love, and
  Aspuru-Guzik]{Aspuru:13}
Whitfield,~J.~D.; Love,~P.~J.; Aspuru-Guzik,~A. \emph{Phys.\ Chem.\ Chem.\
  Phys.} \textbf{2013}, \emph{15}, 397\relax
\mciteBstWouldAddEndPuncttrue
\mciteSetBstMidEndSepPunct{\mcitedefaultmidpunct}
{\mcitedefaultendpunct}{\mcitedefaultseppunct}\relax
\EndOfBibitem
\bibitem[Szalay \latin{et~al.}(2012)Szalay, Muller, Gidofalvi, Lischka, and
  Shepard]{Shepard12}
Szalay,~P.~G.; Muller,~T.; Gidofalvi,~G.; Lischka,~H.; Shepard,~R. \emph{Chem.\
  Rev.} \textbf{2012}, \emph{112}, 108\relax
\mciteBstWouldAddEndPuncttrue
\mciteSetBstMidEndSepPunct{\mcitedefaultmidpunct}
{\mcitedefaultendpunct}{\mcitedefaultseppunct}\relax
\EndOfBibitem
\bibitem[Lyakh \latin{et~al.}(2012)Lyakh, Nusial, Lotrich, and
  Barlett]{Barlett12}
Lyakh,~D.~I.; Nusial,~M.; Lotrich,~V.~F.; Barlett,~R.~J. \emph{Chem.\ Rev.}
  \textbf{2012}, \emph{112}, 182\relax
\mciteBstWouldAddEndPuncttrue
\mciteSetBstMidEndSepPunct{\mcitedefaultmidpunct}
{\mcitedefaultendpunct}{\mcitedefaultseppunct}\relax
\EndOfBibitem
\bibitem[Szabo and Ostlund(1996)Szabo, and Ostlund]{Szabo96}
Szabo,~A.; Ostlund,~N.~S. \emph{Modern Quantum Chemistry: Introduction to
  Advanced Electronic Structure Theory}; Dover Publications, Inc., Mineola,
  1996\relax
\mciteBstWouldAddEndPuncttrue
\mciteSetBstMidEndSepPunct{\mcitedefaultmidpunct}
{\mcitedefaultendpunct}{\mcitedefaultseppunct}\relax
\EndOfBibitem
\bibitem[Helgaker \latin{et~al.}(2014)Helgaker, Jorgensen, and Olsen]{Olsen14}
Helgaker,~T.; Jorgensen,~P.; Olsen,~J. \emph{Molecular Electronic-Structure
  Theory}; John Wiley \& Sons, 2014\relax
\mciteBstWouldAddEndPuncttrue
\mciteSetBstMidEndSepPunct{\mcitedefaultmidpunct}
{\mcitedefaultendpunct}{\mcitedefaultseppunct}\relax
\EndOfBibitem
\bibitem[Martin(2004)]{Martin04}
Martin,~R.~M. \emph{Electronic Structure: Basic Theory and Practical Methods};
  Cambridge University Press, 2004\relax
\mciteBstWouldAddEndPuncttrue
\mciteSetBstMidEndSepPunct{\mcitedefaultmidpunct}
{\mcitedefaultendpunct}{\mcitedefaultseppunct}\relax
\EndOfBibitem
\bibitem[Mardirossian and Head-Gordon(2017)Mardirossian, and
  Head-Gordon]{Head17}
Mardirossian,~N.; Head-Gordon,~M. \emph{Molecular Physics} \textbf{2017},
  \emph{115}, 2315\relax
\mciteBstWouldAddEndPuncttrue
\mciteSetBstMidEndSepPunct{\mcitedefaultmidpunct}
{\mcitedefaultendpunct}{\mcitedefaultseppunct}\relax
\EndOfBibitem
\bibitem[Perdew \latin{et~al.}(2008)Perdew, Ruzsinszky, Csonka, Vydrov,
  Scuseria, Constantin, Zhou, and Burke]{Burke08}
Perdew,~J.~P.; Ruzsinszky,~A.; Csonka,~G.~I.; Vydrov,~O.~A.; Scuseria,~G.~E.;
  Constantin,~L.~A.; Zhou,~X.; Burke,~K. \emph{Phys.\ Rev.\ Lett.}
  \textbf{2008}, \emph{100}, 136406\relax
\mciteBstWouldAddEndPuncttrue
\mciteSetBstMidEndSepPunct{\mcitedefaultmidpunct}
{\mcitedefaultendpunct}{\mcitedefaultseppunct}\relax
\EndOfBibitem
\bibitem[Polo and Cremer(2002)Polo, and Cremer]{Cremer02}
Polo,~V.; Cremer,~E. \emph{Molecular Physics} \textbf{2002}, \emph{100},
  1771\relax
\mciteBstWouldAddEndPuncttrue
\mciteSetBstMidEndSepPunct{\mcitedefaultmidpunct}
{\mcitedefaultendpunct}{\mcitedefaultseppunct}\relax
\EndOfBibitem
\bibitem[Kulik \latin{et~al.}(2006)Kulik, Cococcioni, Scherlis, and
  Marzari]{Marzari06}
Kulik,~H.~J.; Cococcioni,~M.; Scherlis,~D.~A.; Marzari,~N. \emph{Phys. Rev.
  Let.} \textbf{2006}, \emph{97}, 103001\relax
\mciteBstWouldAddEndPuncttrue
\mciteSetBstMidEndSepPunct{\mcitedefaultmidpunct}
{\mcitedefaultendpunct}{\mcitedefaultseppunct}\relax
\EndOfBibitem
\bibitem[Goerigk \latin{et~al.}(2017)Goerigk, Hansen, Bauer, Ehrlich, Najibi,
  and Grimme]{Grimme17}
Goerigk,~L.; Hansen,~A.; Bauer,~C.; Ehrlich,~S.; Najibi,~A.; Grimme,~S.
  \emph{Phys.\ Chem.\ Chem.\ Phys.} \textbf{2017}, \emph{19}, 32184\relax
\mciteBstWouldAddEndPuncttrue
\mciteSetBstMidEndSepPunct{\mcitedefaultmidpunct}
{\mcitedefaultendpunct}{\mcitedefaultseppunct}\relax
\EndOfBibitem
\bibitem[Kirklin1 \latin{et~al.}(2015)Kirklin1, Saal1, Meredig1, Thompson1,
  Doak, Aykol, Ruhl, and Wolverton]{Wolverton15}
Kirklin1,~S.; Saal1,~J.~E.; Meredig1,~B.; Thompson1,~A.; Doak,~J.~W.;
  Aykol,~M.; Ruhl,~S.; Wolverton,~C. \emph{npj Comp. Mater.} \textbf{2015},
  \emph{1}, 15010\relax
\mciteBstWouldAddEndPuncttrue
\mciteSetBstMidEndSepPunct{\mcitedefaultmidpunct}
{\mcitedefaultendpunct}{\mcitedefaultseppunct}\relax
\EndOfBibitem
\bibitem[Grimme(2006)]{Grimme06}
Grimme,~S. \emph{Angew. Chem. Int. Ed.} \textbf{2006}, \emph{45}, 4460\relax
\mciteBstWouldAddEndPuncttrue
\mciteSetBstMidEndSepPunct{\mcitedefaultmidpunct}
{\mcitedefaultendpunct}{\mcitedefaultseppunct}\relax
\EndOfBibitem
\bibitem[Cohen \latin{et~al.}(2008)Cohen, Mori-Sanchez, and Yang]{Yang08}
Cohen,~A.~J.; Mori-Sanchez,~P.; Yang,~W. \emph{Science} \textbf{2008},
  \emph{321}, 792\relax
\mciteBstWouldAddEndPuncttrue
\mciteSetBstMidEndSepPunct{\mcitedefaultmidpunct}
{\mcitedefaultendpunct}{\mcitedefaultseppunct}\relax
\EndOfBibitem
\bibitem[Iori \latin{et~al.}(2012)Iori, Gatti, and Rubio]{Rubio12}
Iori,~F.; Gatti,~M.; Rubio,~A. \emph{Phys.\ Rev.\ B} \textbf{2012}, \emph{85},
  115129\relax
\mciteBstWouldAddEndPuncttrue
\mciteSetBstMidEndSepPunct{\mcitedefaultmidpunct}
{\mcitedefaultendpunct}{\mcitedefaultseppunct}\relax
\EndOfBibitem
\bibitem[Nekrasova \latin{et~al.}(2013)Nekrasova, Pavlova, and
  Sadovskii]{Sadovskii13}
Nekrasova,~I.~A.; Pavlova,~N.~S.; Sadovskii,~M.~V. \emph{Journal of
  Experimental and Theoretical Physics} \textbf{2013}, \emph{116}, 620\relax
\mciteBstWouldAddEndPuncttrue
\mciteSetBstMidEndSepPunct{\mcitedefaultmidpunct}
{\mcitedefaultendpunct}{\mcitedefaultseppunct}\relax
\EndOfBibitem
\bibitem[Gerosa \latin{et~al.}(2015)Gerosa, Bottani, Caramella, Onida,
  Valentin, and Pacchioni]{Pacchioni15}
Gerosa,~M.; Bottani,~C.~E.; Caramella,~L.; Onida,~G.; Valentin,~C.~D.;
  Pacchioni,~G. \emph{Phys.\ Rev.\ B} \textbf{2015}, \emph{91}, 155201\relax
\mciteBstWouldAddEndPuncttrue
\mciteSetBstMidEndSepPunct{\mcitedefaultmidpunct}
{\mcitedefaultendpunct}{\mcitedefaultseppunct}\relax
\EndOfBibitem
\bibitem[Hasnip \latin{et~al.}(2014)Hasnip, Refson, Probert, Yates, Clark, and
  Pickard]{Pickard14}
Hasnip,~P.~J.; Refson,~K.; Probert,~M. I.~J.; Yates,~J.~R.; Clark,~S.~J.;
  Pickard,~C.~J. \emph{Phil. Trans. R. Soc. A} \textbf{2014}, \emph{372},
  20130270\relax
\mciteBstWouldAddEndPuncttrue
\mciteSetBstMidEndSepPunct{\mcitedefaultmidpunct}
{\mcitedefaultendpunct}{\mcitedefaultseppunct}\relax
\EndOfBibitem
\bibitem[Harrison(2000)]{Harrison00}
Harrison,~J.~F. \emph{Chem. Rev.} \textbf{2000}, \emph{100}, 679\relax
\mciteBstWouldAddEndPuncttrue
\mciteSetBstMidEndSepPunct{\mcitedefaultmidpunct}
{\mcitedefaultendpunct}{\mcitedefaultseppunct}\relax
\EndOfBibitem
\bibitem[Jensen \latin{et~al.}(2007)Jensen, Roos, and Ryde]{Ryde07}
Jensen,~K.~P.; Roos,~B.~O.; Ryde,~U. \emph{J. Chem. Phys.} \textbf{2007},
  \emph{126}, 014103\relax
\mciteBstWouldAddEndPuncttrue
\mciteSetBstMidEndSepPunct{\mcitedefaultmidpunct}
{\mcitedefaultendpunct}{\mcitedefaultseppunct}\relax
\EndOfBibitem
\bibitem[Chakraborty \latin{et~al.}(2018)Chakraborty, Dixit, Aurbach, and
  Major]{Major18}
Chakraborty,~A.; Dixit,~M.; Aurbach,~D.; Major,~D.~T. \emph{npj Computational
  materials} \textbf{2018}, \emph{60}\relax
\mciteBstWouldAddEndPuncttrue
\mciteSetBstMidEndSepPunct{\mcitedefaultmidpunct}
{\mcitedefaultendpunct}{\mcitedefaultseppunct}\relax
\EndOfBibitem
\bibitem[Perdew(1985)]{Perdew85}
Perdew,~P.~J. \emph{Int. J. Quantum Chem.} \textbf{1985}, \emph{28}, 497\relax
\mciteBstWouldAddEndPuncttrue
\mciteSetBstMidEndSepPunct{\mcitedefaultmidpunct}
{\mcitedefaultendpunct}{\mcitedefaultseppunct}\relax
\EndOfBibitem
\bibitem[Kong \latin{et~al.}(2015)Kong, Longo, Park, Y., Yeon, Park, Wang, KC,
  Doo, and Cho]{Cho15}
Kong,~F.; Longo,~R.~C.; Park,~M.-S.; Y.,~J.; Yeon,~D.-H.; Park,~J.-H.;
  Wang,~W.-H.; KC,~S.; Doo,~S.-G.; Cho,~K. \emph{J. Mater. Chem. A}
  \textbf{2015}, \emph{3}, 8489\relax
\mciteBstWouldAddEndPuncttrue
\mciteSetBstMidEndSepPunct{\mcitedefaultmidpunct}
{\mcitedefaultendpunct}{\mcitedefaultseppunct}\relax
\EndOfBibitem
\bibitem[Chevrier \latin{et~al.}(2010)Chevrier, Ong, Armiento, Chan, and
  Ceder]{Ceder10}
Chevrier,~V.~L.; Ong,~S.~P.; Armiento,~R.; Chan,~M. K.~Y.; Ceder,~G.
  \emph{Phys.\ Rev.\ B} \textbf{2010}, \emph{82}, 075122\relax
\mciteBstWouldAddEndPuncttrue
\mciteSetBstMidEndSepPunct{\mcitedefaultmidpunct}
{\mcitedefaultendpunct}{\mcitedefaultseppunct}\relax
\EndOfBibitem
\bibitem[Richards \latin{et~al.}(2016)Richards, Miara, Wang, Kim, and
  Ceder]{Ceder16a}
Richards,~W.~D.; Miara,~L.~J.; Wang,~Y.; Kim,~J.~C.; Ceder,~G. \emph{Chem.\
  Mater.} \textbf{2016}, \emph{28}, 266\relax
\mciteBstWouldAddEndPuncttrue
\mciteSetBstMidEndSepPunct{\mcitedefaultmidpunct}
{\mcitedefaultendpunct}{\mcitedefaultseppunct}\relax
\EndOfBibitem
\bibitem[Bruce \latin{et~al.}(1999)Bruce, Armstrong, and
  Gitzendanner]{Gitzendanner99}
Bruce,~P.~G.; Armstrong,~A.~R.; Gitzendanner,~R.~L. \emph{J. Mater. Chem.}
  \textbf{1999}, \emph{9}, 193\relax
\mciteBstWouldAddEndPuncttrue
\mciteSetBstMidEndSepPunct{\mcitedefaultmidpunct}
{\mcitedefaultendpunct}{\mcitedefaultseppunct}\relax
\EndOfBibitem
\bibitem[Cococcioni and de~Gironcoli(2005)Cococcioni, and
  de~Gironcoli]{Gironcoli05}
Cococcioni,~M.; de~Gironcoli,~S. \emph{Phys. Rev. B} \textbf{2005}, \emph{71},
  035105\relax
\mciteBstWouldAddEndPuncttrue
\mciteSetBstMidEndSepPunct{\mcitedefaultmidpunct}
{\mcitedefaultendpunct}{\mcitedefaultseppunct}\relax
\EndOfBibitem
\bibitem[Kulik and Marzari(2010)Kulik, and Marzari]{Marzari10}
Kulik,~H.~J.; Marzari,~N. \emph{J. Chem. Phys.} \textbf{2010}, \emph{133},
  114103\relax
\mciteBstWouldAddEndPuncttrue
\mciteSetBstMidEndSepPunct{\mcitedefaultmidpunct}
{\mcitedefaultendpunct}{\mcitedefaultseppunct}\relax
\EndOfBibitem
\bibitem[Zhou \latin{et~al.}(2004)Zhou, Cococcioni, Marianetti, Morgan, and
  Ceder]{Ceder04}
Zhou,~F.; Cococcioni,~M.; Marianetti,~C.; Morgan,~D.; Ceder,~G. \emph{Phys.\
  Rev.\ B} \textbf{2004}, \emph{70}, 235121\relax
\mciteBstWouldAddEndPuncttrue
\mciteSetBstMidEndSepPunct{\mcitedefaultmidpunct}
{\mcitedefaultendpunct}{\mcitedefaultseppunct}\relax
\EndOfBibitem
\bibitem[Adamo and Barone(1999)Adamo, and Barone]{Barone99}
Adamo,~C.; Barone,~V. \emph{J. Phys. Chem. C} \textbf{1999}, \emph{110},
  6158\relax
\mciteBstWouldAddEndPuncttrue
\mciteSetBstMidEndSepPunct{\mcitedefaultmidpunct}
{\mcitedefaultendpunct}{\mcitedefaultseppunct}\relax
\EndOfBibitem
\bibitem[Andrew and Bell(2013)Andrew, and Bell]{Bell13}
Andrew,~G.; Bell,~A.~T. \emph{J. Phys. Chem. C} \textbf{2013}, \emph{117},
  25562\relax
\mciteBstWouldAddEndPuncttrue
\mciteSetBstMidEndSepPunct{\mcitedefaultmidpunct}
{\mcitedefaultendpunct}{\mcitedefaultseppunct}\relax
\EndOfBibitem
\bibitem[Barden \latin{et~al.}(1999)Barden, Rienstra-Kiracofe, and
  Schaefer]{Schaefer99}
Barden,~C.~J.; Rienstra-Kiracofe,~J.~C.; Schaefer,~H.~F. \emph{J. Chem. Phys.}
  \textbf{1999}, \emph{113}, 690\relax
\mciteBstWouldAddEndPuncttrue
\mciteSetBstMidEndSepPunct{\mcitedefaultmidpunct}
{\mcitedefaultendpunct}{\mcitedefaultseppunct}\relax
\EndOfBibitem
\bibitem[Janthon \latin{et~al.}(2014)Janthon, Luo, Kozlov, Vines, Limtrakul,
  d.~G.~Truhlar, and Illas]{Illas14}
Janthon,~P.; Luo,~S.; Kozlov,~S.~M.; Vines,~F.; Limtrakul,~J.; d.~G.~Truhlar,;
  Illas,~F. \emph{J. Chem. Theory Comput.} \textbf{2014}, \emph{10}, 3832\relax
\mciteBstWouldAddEndPuncttrue
\mciteSetBstMidEndSepPunct{\mcitedefaultmidpunct}
{\mcitedefaultendpunct}{\mcitedefaultseppunct}\relax
\EndOfBibitem
\bibitem[Chai and Head-Gordon(2008)Chai, and Head-Gordon]{Gordon08}
Chai,~J.-D.; Head-Gordon,~M. \emph{J. Chem. Phys.} \textbf{2008}, \emph{128},
  084106\relax
\mciteBstWouldAddEndPuncttrue
\mciteSetBstMidEndSepPunct{\mcitedefaultmidpunct}
{\mcitedefaultendpunct}{\mcitedefaultseppunct}\relax
\EndOfBibitem
\bibitem[Heyd \latin{et~al.}(2003)Heyd, Scuseria, and Ernzerhof]{Ernzerhof03}
Heyd,~J.; Scuseria,~G.~E.; Ernzerhof,~M. \emph{J. Chem. Phys.} \textbf{2003},
  \emph{118}, 8207\relax
\mciteBstWouldAddEndPuncttrue
\mciteSetBstMidEndSepPunct{\mcitedefaultmidpunct}
{\mcitedefaultendpunct}{\mcitedefaultseppunct}\relax
\EndOfBibitem
\bibitem[Chai and Head-Gordon(2008)Chai, and Head-Gordon]{Gordon08b}
Chai,~J.-D.; Head-Gordon,~M. \emph{Phys. Chem. Chem. Phys.} \textbf{2008},
  \emph{10}, 6615\relax
\mciteBstWouldAddEndPuncttrue
\mciteSetBstMidEndSepPunct{\mcitedefaultmidpunct}
{\mcitedefaultendpunct}{\mcitedefaultseppunct}\relax
\EndOfBibitem
\bibitem[Cococcioni and Marzari(2019)Cococcioni, and Marzari]{Marzari19}
Cococcioni,~M.; Marzari,~N. \emph{Phys. Rev. Materials} \textbf{2019},
  \emph{3}, 033801\relax
\mciteBstWouldAddEndPuncttrue
\mciteSetBstMidEndSepPunct{\mcitedefaultmidpunct}
{\mcitedefaultendpunct}{\mcitedefaultseppunct}\relax
\EndOfBibitem
\bibitem[Klimes \latin{et~al.}(2010)Klimes, Bowler, and
  Michaelides]{Michaelides10}
Klimes,~J.; Bowler,~D.~R.; Michaelides,~A. \emph{J. Phys.: Condens. Matter}
  \textbf{2010}, \emph{22}, 022201\relax
\mciteBstWouldAddEndPuncttrue
\mciteSetBstMidEndSepPunct{\mcitedefaultmidpunct}
{\mcitedefaultendpunct}{\mcitedefaultseppunct}\relax
\EndOfBibitem
\bibitem[Nielsen and Chunag(2020)Nielsen, and Chunag]{Nielsen10}
Nielsen,~M.~A.; Chunag,~I.~L. \emph{Quantum Computational and Quantum
  Information}; Cambridge University Press, 2020\relax
\mciteBstWouldAddEndPuncttrue
\mciteSetBstMidEndSepPunct{\mcitedefaultmidpunct}
{\mcitedefaultendpunct}{\mcitedefaultseppunct}\relax
\EndOfBibitem
\bibitem[Cao \latin{et~al.}(2019)Cao, Romero, Olson, Degroote, Johnson,
  Kieferova, Kivlichan, Menke, Peropadre, Sawaya, Sim, Veis, and
  Aspuru-Guzik]{Aspuru19}
Cao,~Y.; Romero,~J.; Olson,~J.~P.; Degroote,~M.; Johnson,~P.~D.; Kieferova,~M.;
  Kivlichan,~I.~D.; Menke,~T.; Peropadre,~B.; Sawaya,~N. P.~D.; Sim,~S.;
  Veis,~L.; Aspuru-Guzik,~A. \emph{Chem.\ Rev.} \textbf{2019}, \emph{119},
  10856\relax
\mciteBstWouldAddEndPuncttrue
\mciteSetBstMidEndSepPunct{\mcitedefaultmidpunct}
{\mcitedefaultendpunct}{\mcitedefaultseppunct}\relax
\EndOfBibitem
\bibitem[McArdle \latin{et~al.}(2020)McArdle, Endo, Aspuru-Guzik, Benjamin, and
  Yuan]{Aspuru:20}
McArdle,~S.; Endo,~S.; Aspuru-Guzik,~A.; Benjamin,~S.~C.; Yuan,~X. \emph{Review
  of Modern Physics} \textbf{2020}, \emph{92}, 015003\relax
\mciteBstWouldAddEndPuncttrue
\mciteSetBstMidEndSepPunct{\mcitedefaultmidpunct}
{\mcitedefaultendpunct}{\mcitedefaultseppunct}\relax
\EndOfBibitem
\bibitem[Head-Marsden \latin{et~al.}(2021)Head-Marsden, Flick, Ciccarino, and
  Narang]{Narang21}
Head-Marsden,~K.; Flick,~J.; Ciccarino,~C.~J.; Narang,~P. \emph{Chem.\ Rev.}
  \textbf{2021}, \emph{121}, 3061\relax
\mciteBstWouldAddEndPuncttrue
\mciteSetBstMidEndSepPunct{\mcitedefaultmidpunct}
{\mcitedefaultendpunct}{\mcitedefaultseppunct}\relax
\EndOfBibitem
\bibitem[Bauer \latin{et~al.}(2020)Bauer, Bravyi, Motta, and Chan]{Chan20}
Bauer,~B.; Bravyi,~S.; Motta,~M.; Chan,~G. K.-L. \emph{Chem.\ Rev.}
  \textbf{2020}, \emph{120}, 12685\relax
\mciteBstWouldAddEndPuncttrue
\mciteSetBstMidEndSepPunct{\mcitedefaultmidpunct}
{\mcitedefaultendpunct}{\mcitedefaultseppunct}\relax
\EndOfBibitem
\bibitem[Aspuru-Guzik \latin{et~al.}(2005)Aspuru-Guzik, Dutoi, Love, and
  Head-Gordon]{Gordon05}
Aspuru-Guzik,~A.; Dutoi,~A.~D.; Love,~P.~J.; Head-Gordon,~M. \emph{Science}
  \textbf{2005}, \emph{309}, 1704\relax
\mciteBstWouldAddEndPuncttrue
\mciteSetBstMidEndSepPunct{\mcitedefaultmidpunct}
{\mcitedefaultendpunct}{\mcitedefaultseppunct}\relax
\EndOfBibitem
\bibitem[Bharit \latin{et~al.}(2022)Bharit, Cervera-Lierta, Kyaw, Huag,
  Alperin-Lea, Anand, Degroot, Heimonen, Kottmann, Menke, Mok, Sim, Kwek, and
  Aspuru-Guzik]{Aspuru:22}
Bharit,~K.; Cervera-Lierta,~A.; Kyaw,~T.~H.; Huag,~T.; Alperin-Lea,~S.;
  Anand,~A.; Degroot,~M.; Heimonen,~H.; Kottmann,~J.~S.; Menke,~T.; Mok,~W.-K.;
  Sim,~S.; Kwek,~L.-C.; Aspuru-Guzik,~A. \emph{Review of Modern Physics}
  \textbf{2022}, \emph{94}, 015004\relax
\mciteBstWouldAddEndPuncttrue
\mciteSetBstMidEndSepPunct{\mcitedefaultmidpunct}
{\mcitedefaultendpunct}{\mcitedefaultseppunct}\relax
\EndOfBibitem
\bibitem[Delgado \latin{et~al.}(2022)Delgado, Casares, Reis, Zini, Campos,
  Cruz-Hernandez, Voigt, Lowe, Jahangiri, Martin-Delgado, Muller, and
  Arrazola]{Arrazola22}
Delgado,~A.; Casares,~P. A.~M.; Reis,~R.~D.; Zini,~M.~S.; Campos,~R.;
  Cruz-Hernandez,~N.; Voigt,~A.-C.; Lowe,~A.; Jahangiri,~S.;
  Martin-Delgado,~M.~A.; Muller,~J.~E.; Arrazola,~J.~M. arXiv:2204.11890,
  2022\relax
\mciteBstWouldAddEndPuncttrue
\mciteSetBstMidEndSepPunct{\mcitedefaultmidpunct}
{\mcitedefaultendpunct}{\mcitedefaultseppunct}\relax
\EndOfBibitem
\bibitem[Kim \latin{et~al.}(2022)Kim, Liu, Pallister, Pol, and
  Roberts]{Roberts22}
Kim,~I.~H.; Liu,~Y.-H.; Pallister,~S.; Pol,~W.; Roberts,~S. \emph{Physical
  Review Research} \textbf{2022}, \emph{4}, 023019\relax
\mciteBstWouldAddEndPuncttrue
\mciteSetBstMidEndSepPunct{\mcitedefaultmidpunct}
{\mcitedefaultendpunct}{\mcitedefaultseppunct}\relax
\EndOfBibitem
\bibitem[McClean \latin{et~al.}(2016)McClean, Romero, Babbush, and
  Aspuru-Guzik]{Aspuru:16}
McClean,~J.~R.; Romero,~J.; Babbush,~R.; Aspuru-Guzik,~A. \emph{New \ J.\
  Phys.} \textbf{2016}, \emph{18}, 023023\relax
\mciteBstWouldAddEndPuncttrue
\mciteSetBstMidEndSepPunct{\mcitedefaultmidpunct}
{\mcitedefaultendpunct}{\mcitedefaultseppunct}\relax
\EndOfBibitem
\bibitem[Anand \latin{et~al.}(2022)Anand, Schleich, Alperin-Lea, Jensen, Sim,
  Diaz-Tinoco, Kottmann, Degroote, Izmaylov, and Aspuru-Guzik]{Izmaylov22}
Anand,~A.; Schleich,~P.; Alperin-Lea,~S.; Jensen,~P. W.~K.; Sim,~S.;
  Diaz-Tinoco,~M.; Kottmann,~J.~S.; Degroote,~M.; Izmaylov,~A.~F.;
  Aspuru-Guzik,~A. \emph{Chem.\ Soc.\ Rev.} \textbf{2022}, \emph{51},
  1659\relax
\mciteBstWouldAddEndPuncttrue
\mciteSetBstMidEndSepPunct{\mcitedefaultmidpunct}
{\mcitedefaultendpunct}{\mcitedefaultseppunct}\relax
\EndOfBibitem
\bibitem[Romero \latin{et~al.}(2019)Romero, Babbush, McClean, Hempel, Love, and
  Aspuru-Guzik]{Babbush19}
Romero,~J.; Babbush,~R.; McClean,~J.~R.; Hempel,~C.; Love,~P.; Aspuru-Guzik,~A.
  \emph{Quantum Sci.\ Technol.} \textbf{2019}, \emph{4}, 014008\relax
\mciteBstWouldAddEndPuncttrue
\mciteSetBstMidEndSepPunct{\mcitedefaultmidpunct}
{\mcitedefaultendpunct}{\mcitedefaultseppunct}\relax
\EndOfBibitem
\bibitem[Lee \latin{et~al.}(2019)Lee, Huggins, Head-gordon, and
  Whaley]{Whaley19}
Lee,~J.; Huggins,~W.~J.; Head-gordon,~M.; Whaley,~K.~B. \emph{J.\ Chem.\
  theory\ Comput.} \textbf{2019}, \emph{15}, 311\relax
\mciteBstWouldAddEndPuncttrue
\mciteSetBstMidEndSepPunct{\mcitedefaultmidpunct}
{\mcitedefaultendpunct}{\mcitedefaultseppunct}\relax
\EndOfBibitem
\bibitem[Huron \latin{et~al.}(1973)Huron, Malrieu, and Rancurel]{Rancurel73}
Huron,~B.; Malrieu,~J.~P.; Rancurel,~P. \emph{J.\ Chem.\ Phys.} \textbf{1973},
  \emph{58}, 5745\relax
\mciteBstWouldAddEndPuncttrue
\mciteSetBstMidEndSepPunct{\mcitedefaultmidpunct}
{\mcitedefaultendpunct}{\mcitedefaultseppunct}\relax
\EndOfBibitem
\bibitem[Dyall(1994)]{Dyall94}
Dyall,~K.~G. \emph{J.\ Chem.\ Phys.} \textbf{1994}, \emph{102}, 4909\relax
\mciteBstWouldAddEndPuncttrue
\mciteSetBstMidEndSepPunct{\mcitedefaultmidpunct}
{\mcitedefaultendpunct}{\mcitedefaultseppunct}\relax
\EndOfBibitem
\bibitem[Not()]{Note-1}
https://materialsproject.org\relax
\mciteBstWouldAddEndPuncttrue
\mciteSetBstMidEndSepPunct{\mcitedefaultmidpunct}
{\mcitedefaultendpunct}{\mcitedefaultseppunct}\relax
\EndOfBibitem
\bibitem[Not()]{Note-2}
http://jmol.sourceforge.net\relax
\mciteBstWouldAddEndPuncttrue
\mciteSetBstMidEndSepPunct{\mcitedefaultmidpunct}
{\mcitedefaultendpunct}{\mcitedefaultseppunct}\relax
\EndOfBibitem
\bibitem[Not()]{Note-3}
http://jmol.sourceforge.net\relax
\mciteBstWouldAddEndPuncttrue
\mciteSetBstMidEndSepPunct{\mcitedefaultmidpunct}
{\mcitedefaultendpunct}{\mcitedefaultseppunct}\relax
\EndOfBibitem
\bibitem[Balabanova and Peterson(2006)Balabanova, and Peterson]{Peterson06}
Balabanova,~N.~B.; Peterson,~K.~A. \emph{J. Chem. Phys.} \textbf{2006},
  \emph{125}, 074110\relax
\mciteBstWouldAddEndPuncttrue
\mciteSetBstMidEndSepPunct{\mcitedefaultmidpunct}
{\mcitedefaultendpunct}{\mcitedefaultseppunct}\relax
\EndOfBibitem
\bibitem[Sun \latin{et~al.}(2020)Sun, Zhang, Banerjee, Bao, Barbry, Blunt,
  Bogdanov, Booth, Chen, Cui, Eriksen, Y.~Gao, Hermann, Hermes, Koh, Koval,
  Lehtola, Li, Liu, Mardirossian, McClain, Motta, Mussard, Pham, Pulkin,
  Purwanto, Robinson, Ronca, Sayfutyarova, Scheurer, Schurkus, Smith, Sun, Sun,
  Upadhyay, Wagner, Wang, White, Whitfield, Williamson, Wouters, Yang, Yu, Zhu,
  Berkelbach, Sharma, Sokolov, , and Chan]{Berkelbach20}
Sun,~Q. \latin{et~al.}  \emph{J.\ chem.\ Phys.} \textbf{2020}, \emph{153},
  024109\relax
\mciteBstWouldAddEndPuncttrue
\mciteSetBstMidEndSepPunct{\mcitedefaultmidpunct}
{\mcitedefaultendpunct}{\mcitedefaultseppunct}\relax
\EndOfBibitem
\bibitem[Sun \latin{et~al.}(2018)Sun, Berkelbach, Blunt, Booth, Guo, Li, Liu,
  McClain, Sharma, Wouters, , and Chan]{Berkelbach18}
Sun,~Q.; Berkelbach,~T.~C.; Blunt,~N.~S.; Booth,~G.~H.; Guo,~S.; Li,~Z.;
  Liu,~J.; McClain,~J.; Sharma,~S.; Wouters,~S.; ; Chan,~G. K.-L. \emph{WIREs\
  Comput.\ Mol.\ Sci.} \textbf{2018}, \emph{8}, e1340\relax
\mciteBstWouldAddEndPuncttrue
\mciteSetBstMidEndSepPunct{\mcitedefaultmidpunct}
{\mcitedefaultendpunct}{\mcitedefaultseppunct}\relax
\EndOfBibitem
\bibitem[Sun(2015)]{Sun15}
Sun,~Q. \emph{J. Comp. Chem.} \textbf{2015}, \emph{36}, 1664\relax
\mciteBstWouldAddEndPuncttrue
\mciteSetBstMidEndSepPunct{\mcitedefaultmidpunct}
{\mcitedefaultendpunct}{\mcitedefaultseppunct}\relax
\EndOfBibitem
\bibitem[Laubach \latin{et~al.}(2009)Laubach, Laubach, Schemidt, Ensling,
  Schmid, Jaegermann, Thiben, Nikolowski, and Ehrenberg]{Ehrenberg09}
Laubach,~S.; Laubach,~S.; Schemidt,~P.~C.; Ensling,~D.; Schmid,~S.;
  Jaegermann,~W.; Thiben,~A.; Nikolowski,~K.; Ehrenberg,~H. \emph{Phys.\ Chem.\
  Chem.\ Phys.} \textbf{2009}, \emph{11}, 3278\relax
\mciteBstWouldAddEndPuncttrue
\mciteSetBstMidEndSepPunct{\mcitedefaultmidpunct}
{\mcitedefaultendpunct}{\mcitedefaultseppunct}\relax
\EndOfBibitem
\bibitem[van Elp \latin{et~al.}(1991)van Elp, Wieland, Eske, Kuiper, and
  Sawatzky]{Sawatzky91}
van Elp,~J.; Wieland,~J.~L.; Eske,~H.; Kuiper,~P.; Sawatzky,~G.~A. \emph{Phys.\
  Rev.\ B} \textbf{1991}, \emph{44}, 6090\relax
\mciteBstWouldAddEndPuncttrue
\mciteSetBstMidEndSepPunct{\mcitedefaultmidpunct}
{\mcitedefaultendpunct}{\mcitedefaultseppunct}\relax
\EndOfBibitem
\bibitem[Ensling \latin{et~al.}(2010)Ensling, Thissen, Laubach, Schmidt, and
  Jaegermann]{Jaegermann10}
Ensling,~D.; Thissen,~A.; Laubach,~S.; Schmidt,~P.~C.; Jaegermann,~W.
  \emph{Phys.\ Rev. B} \textbf{2010}, \emph{82}, 195431\relax
\mciteBstWouldAddEndPuncttrue
\mciteSetBstMidEndSepPunct{\mcitedefaultmidpunct}
{\mcitedefaultendpunct}{\mcitedefaultseppunct}\relax
\EndOfBibitem
\bibitem[Not()]{Note-4}
A. Tranter, C. Di Paola, D. M. Ramo, D. Z. Manrique, D. Gowland, G.
  Greene-Diniz, G. Christopoulou, I. Polyak, I. Khan, J. Pilipczuk, J. Kirsopp,
  K. Yamamoto, M. Tudorovskaya, M. Krompiec, N. Fitzpatrick, InQuanto -
  Introduction to the InQuanto Computational Chemistry Platform For Quantum
  Computers, 2022,
  https://medium.com/cambridge-quantum-computing/4fced08d66cc\relax
\mciteBstWouldAddEndPuncttrue
\mciteSetBstMidEndSepPunct{\mcitedefaultmidpunct}
{\mcitedefaultendpunct}{\mcitedefaultseppunct}\relax
\EndOfBibitem
\bibitem[Seeley \latin{et~al.}(2012)Seeley, Richard, and Love]{Love12}
Seeley,~J.~T.; Richard,~M.~J.; Love,~P.~J. \emph{J.\ Chem.\ Phys.}
  \textbf{2012}, \emph{137}, 224109\relax
\mciteBstWouldAddEndPuncttrue
\mciteSetBstMidEndSepPunct{\mcitedefaultmidpunct}
{\mcitedefaultendpunct}{\mcitedefaultseppunct}\relax
\EndOfBibitem
\bibitem[Not()]{Note-5}
https://quantum-computing.ibm.com/lab/docs/iql/manage/simulator/\relax
\mciteBstWouldAddEndPuncttrue
\mciteSetBstMidEndSepPunct{\mcitedefaultmidpunct}
{\mcitedefaultendpunct}{\mcitedefaultseppunct}\relax
\EndOfBibitem
\bibitem[Hughes1 \latin{et~al.}(2008)Hughes1, Dane, Ernst, Hergert, Luders,
  Staunton, Szotek, and Temmerman]{Temmerman08}
Hughes1,~I.~D.; Dane,~M.; Ernst,~A.; Hergert,~W.; Luders,~M.; Staunton,~J.~B.;
  Szotek,~Z.; Temmerman,~W.~M. \emph{New Journal of Physics} \textbf{2008},
  \emph{10}, 063010\relax
\mciteBstWouldAddEndPuncttrue
\mciteSetBstMidEndSepPunct{\mcitedefaultmidpunct}
{\mcitedefaultendpunct}{\mcitedefaultseppunct}\relax
\EndOfBibitem
\bibitem[Shee \latin{et~al.}(2019)Shee, Rudshteyn, Arthur, Zhang, Reichman, and
  Friesner]{Friesner19}
Shee,~J.; Rudshteyn,~B.; Arthur,~E.~J.; Zhang,~S.; Reichman,~D.~R.;
  Friesner,~R.~A. \emph{J. Chem. Theory Comput.} \textbf{2019}, \emph{15},
  2346\relax
\mciteBstWouldAddEndPuncttrue
\mciteSetBstMidEndSepPunct{\mcitedefaultmidpunct}
{\mcitedefaultendpunct}{\mcitedefaultseppunct}\relax
\EndOfBibitem
\bibitem[Aoto \latin{et~al.}(2017)Aoto, de~L.~Batista, Kohn, and
  de~Oliveira-Filho]{Oliveira17}
Aoto,~Y.~A.; de~L.~Batista,~A.~P.; Kohn,~A.; de~Oliveira-Filho,~A. G.~S.
  \emph{J. Chem. Theory Comput.} \textbf{2017}, \emph{13}, 5291\relax
\mciteBstWouldAddEndPuncttrue
\mciteSetBstMidEndSepPunct{\mcitedefaultmidpunct}
{\mcitedefaultendpunct}{\mcitedefaultseppunct}\relax
\EndOfBibitem
\bibitem[Xu \latin{et~al.}(2015)Xu, Zhang, Tang, and Truhlar]{Truhlar15}
Xu,~X.; Zhang,~W.; Tang,~M.; Truhlar,~D.~G. \emph{J. Chem. Theory Comput.}
  \textbf{2015}, \emph{11}, 2036\relax
\mciteBstWouldAddEndPuncttrue
\mciteSetBstMidEndSepPunct{\mcitedefaultmidpunct}
{\mcitedefaultendpunct}{\mcitedefaultseppunct}\relax
\EndOfBibitem
\bibitem[Hait \latin{et~al.}(2019)Hait, Tubman, Levine, Whaley, and
  Head-Gordon]{Gordon19}
Hait,~D.; Tubman,~N.~M.; Levine,~D.~S.; Whaley,~K.~B.; Head-Gordon,~M.
  \emph{J.\ Chem.\ Theory\ Comput.} \textbf{2019}, \emph{15}, 5370\relax
\mciteBstWouldAddEndPuncttrue
\mciteSetBstMidEndSepPunct{\mcitedefaultmidpunct}
{\mcitedefaultendpunct}{\mcitedefaultseppunct}\relax
\EndOfBibitem
\bibitem[Bartlett and Musial(2007)Bartlett, and Musial]{Musial07}
Bartlett,~R.~J.; Musial,~M. \emph{Review of Modern Physics} \textbf{2007},
  \emph{79}, 291\relax
\mciteBstWouldAddEndPuncttrue
\mciteSetBstMidEndSepPunct{\mcitedefaultmidpunct}
{\mcitedefaultendpunct}{\mcitedefaultseppunct}\relax
\EndOfBibitem
\end{mcitethebibliography}

\end{document}